\documentclass[usenatbib,useAMS]{mn2e}
\usepackage{graphicx,epstopdf,float,color,hyperref,amsmath,pdflscape,multirow,epstopdf,graphics}

\def\mnras{MNRAS}
\def\apj{ApJ}

\def\apjs{ApJS}
\def\aj{AJ}
\def\araa{Annu. Rev. Astron. Astrophys.}
\def\prd{PRD}
\def\aap{Astron. \& Astrophys.}

\def\physrep{Physics Reports}
\def\aaps{A\&AS}

\newcommand{\corr}{\textcolor{black}}

\title[Weak Lensing Measurements in Simulations of Radio Images]{Weak Lensing Measurements in Simulations of Radio Images}
\author[Patel et al.]{Prina Patel$^{1,2}$\thanks{prina83@gmail.com}, Filipe B. Abdalla$^3$, David J. Bacon$^2$, Barnaby Rowe$^{3,4}$, \newauthor Oleg M. Smirnov$^{5,6}$, Rob J. Beswick$^{7,8}$ \\\\
$^1$Astrophysics, Cosmology Gravity Centre, and Department of Mathematics and Applied Mathematics, \\University of Cape Town, Cape Town, 7701, South Africa.\\
$^2$Institute of Cosmology \& Gravitation, University of Portsmouth, Dennis Sciama Building, Portsmouth, PO1 3FX\\
$^3$Department of Physics and Astronomy, University College London, Gower Street, London WC1E 6BT\\
$^4$Institut d'Astrophysique de Paris, UMR7095 CNRS, Universit\'{e} Pierre et Marie Curie -- Paris 6, 98 bis, \\Boulevard Arago, 75014 Paris, France. \\
$^5$Department of Physics and Electronics, Rhodes University, PO Box 94, Grahamstown, 6140, South Africa\\
$^6$SKA South Africa, 3rd Floor, The Park, Park Road, Pinelands, 7405, South Africa\\
$^7${\it e}-MERLIN/VLBI National Radio Astronomy Facility, Jodrell Bank Observatory, The University of Manchester, \\Macclesfield, Cheshire, SK11 9DL, UK.\\
$^8$Jodrell Bank Centre for Astrophysics, School of Physics and Astronomy, The University of Manchester, Oxford Road,\\ Manchester, M13 9PL, UK}

\begin{document}
\date{Accepted ----. Received ----; in original form ----.}
\pagerange{\pageref{firstpage}--\pageref{lastpage}} \pubyear{2013}
\maketitle
\label{firstpage}

\begin{abstract}
We present a study of weak lensing shear measurements for simulated galaxy images at radio wavelengths. \corr{We construct a simulation pipeline into which we can input galaxy images of known shapelet ellipticity,} and with which we then simulate observations with eMERLIN and the international LOFAR array. The simulations include the effects of the CLEAN algorithm, uv sampling, observing angle, and visibility noise, and produce realistic restored images of the galaxies. \corr{We apply a shapelet-based shear measurement method to these images and test our ability to recover the true source shapelet ellipticities.} We model and deconvolve the effective PSF, and find suitable parameters for CLEAN and shapelet decomposition of galaxies. We demonstrate that ellipticities can be measured faithfully in these radio simulations, with no evidence of an additive bias and a modest (10\%) multiplicative bias on the ellipticity measurements. Our simulation pipeline can be used to test shear measurement procedures and systematics for the next generation of radio telescopes.
\end{abstract}

\section{Introduction}
\label{intro}
The bending of light due to the large scale structure in the Universe is a powerful probe in studying the underlying matter distribution. This is due in part to the gravitational lensing being insensitive to the type of matter causing the light deflection \citep[e.g.][]{2001PhR...340..291B,2003ARA&A..41..645R, 2008PhR...462...67M}. There is the added benefit of lensing being sensitive to the geometry of the Universe, making it useful in the study of dark energy \citep{2002PhRvD..65f3001H}. 

To date, almost all cosmic shear analyses have been conducted at optical wavelengths. However, radio astronomy is currently going through a period of rapid expansion which will make future radio surveys competitive for lensing studies. New radio telescopes will have sensitivities that will reach a level where the radio emission from ordinary galaxies will be routinely resolved (e.g. with eMERLIN\footnote{http://www.e-merlin.ac.uk/}, LOFAR\footnote{http://www.lofar.org/}, and eventually SKA\footnote{http://www.skatelescope.org/}; c.f. \citealp{2004MNRAS.352..131S}); hence radio source densities will become comparable to those found at optical wavelengths. Also, radio interferometers have a well known and deterministic dirty beam pattern, which may be an advantage in deconvolving galaxy shapes for shear measurements.

With the FIRST survey \citep{1995ApJ...450..559B,1997ApJ...475..479W}, \citet{2004ApJ...617..794C} made the first detection of cosmic shear with radio data. This survey has a detection threshold of 1\,mJy, with $\simeq 20$ resolved sources per square degree useable for weak lensing. This is a much lower number density than found in deep optical shear surveys, with a correspondingly lower signal-to-noise on the final cosmological constraints. However, the differential radio source counts at 1.4\,GHz show an increase at flux densities below 1\,mJy, (e.g. \citealp{2004MNRAS.352..131S}), and it is this increase in the number density at the micro-Jansky level that makes future radio weak lensing plausible. 

The feasibility of weak lensing studies at radio wavelengths, and in particular at the micro-Jansky flux levels, was demonstrated in \citet{2010MNRAS.401.2572P}. Due to the low number density of sources used in that work, no significant cosmic signal was detected. However, an upper bound was obtained on a combination of the cosmological parameters $\sigma_{8}$, the normalisation of the matter power spectrum, $\Omega_{m}$, the cosmic matter density parameter, and $z_{m}$, the median redshift of the sources. One of the main conclusions of that work was the need for a detailed study of the systematics involved in radio interferometry and the relevant imaging techniques. 

In this current paper, we pursue that study. We will explore possible systematics which will be important for weak lensing studies with future radio surveys; for example, the systematics which might be introduced by steps such as the CLEAN algorithm or a poor choice of shape measurement parameters. In tandem, we will quantify our current ability to measure realistic galaxy shapes from simulated radio data, increasing our confidence in current and future radio lensing measurements.

The paper is organised as follows: in \S\ref{sims} we describe the whole of our radio image simulation and shape measurement pipeline. We describe the shapelets method which is used throughout this work, and how deconvolution works within its framework and the shape estimator we use. We then describe the shapelets based images that we created as the input for our simulations, before describing how the simulator works to produce realistic radio interferometer images that we use for our analysis. We also describe the telescope configurations we use in this study, and the observational effects that we consider. 

In \S\ref{results} we describe the results of our shape measurements on the simulated images. We assess the appropriate level of CLEANing of images, and examine the modelling of point sources and the effect of changing the position on the sky at which images are observed. We discuss the impact of the shapelet scale parameter in modelling galaxy ellipticities successfully, and the effect of slightly changing the uv sampling. We then present the main result of the paper: how well we are able to recover the input ellipticities with our shape-measurement method. We describe how we add realistic noise into the simulations and what effect this has on our results. In \S\ref{discussion} we end with a discussion of our results, their limitations, and suitable directions for further study.

\section{Simulations of Radio Images}
\label{sims}
The radio image simulation pipeline is built in two parts. We first create a suite of input images, containing sources which constitute the inputs for the main part of the simulation code. These input images are then fed into the simulator that `observes' and `images' them with a given telescope configuration and user defined observation parameters. In this section we start by briefly describing the shapelets technique of \citet{2003MNRAS.338...35R} and \citet{2003MNRAS.338...48R} which we use to create the galaxy shapes in our input images; we will also use shapelet techniques for shape measurement later. We  then describe the creation of the input images before discussing the radio observation pipeline.   

\subsection{Shapelets}
\label{shapelets}
The shapelets description of galaxy shapes is based on decomposing an object's surface brightness $f(\mathbf{x})$ into a series of localised basis functions $B_{n_{1},n_{2}}$, called shapelets:
\begin{equation}
\label{eq:decomp}
f(\mathbf{x})=\sum_{n_{1},n_{2}}f_{n_{1},n_{2}}B_{n_{1},n_{2}}(\mathbf{x};\beta).
\end{equation}  
We briefly describe the relevant parts of the method here and refer the reader to \citet{2003MNRAS.338...35R}, \citet{2003MNRAS.338...48R} and \citet{2005MNRAS.363..197M} for a more detailed description. The basis functions used (in the Cartesian formalism) $B_{n_{1},n_{2}}$, are a localised orthonormal basis set:
\begin{equation}
\label{eq:cartbasis}
B_{n_{1},n_{2}}({\bf x};\beta)=\frac{H_{n_{1}}\left(\frac{x_{1}}{\beta}\right)H_{n_{2}}\left(\frac{x_{2}}{\beta}\right)e^{-\frac{|x|^{2}}{2\beta^{2}}}}{\left(2^{n_{1}n_{2}}\beta^{2}\pi n_{1}!n_{2}! \right) ^{\frac{1}{2}}},
\end{equation}
where $H_{m}(\beta)$ is the $m^{\textrm{th}}$ order Hermite polynomial, and $\beta$ describes the characteristic scale of the basis. We will see that this quantity is very important in our PSF and galaxy shape analysis. $n_{1}+n_{2}$ refers to the order of the basis functions, and in practice all galaxy decompositions are truncated at some order $n_{max}=n_{1}+n_{2}$. $n_{max}$ needs to be  chosen so that the galaxy model is sufficiently detailed to capture ellipticity information. As the basis functions are orthogonal, we can find shapelet coefficients for a galaxy by calculating
\begin{equation}
\label{eq:overlap}
f_{n_{1},n_{2}}=\int\,f(\mathbf{x})B_{n_{1},n_{2}}({\bf x};\beta)\,d^{2}\mathbf{x}.
\end{equation}
\citet{2005MNRAS.363..197M} introduced the closely related polar shapelet formalism (the polar basis set is an orthogonal transformation of the Cartesian set; see \citet{2005MNRAS.363..197M} for details of this transformation), which has a different basis set $P_{n,m}(\theta,\phi;\beta)$, given by
\begin{eqnarray}
P_{n,m}(\theta,\phi;\beta)&=&\frac{(-1)^{(n-|m|)/2}}{\beta^{|m|+1}}\left\{\frac{\left[(n-|m|)/2\right]!}{\pi\left[(n-|m|)/2\right]!}\right\}^{1/2} \nonumber \\ &\times& \theta^{|m|}L^{|m|}_{(n-|m|)/2}\frac{\theta^{2}}{\beta^{2}}e^{-\theta^{2}/2\beta^{2}}e^{-im\phi}.
\end{eqnarray}
In this basis set $\theta$ is the modulus of the complex sky position vector $\theta_{1}+i\theta_{2}$, and $\phi=\arctan(\theta_{2}/\theta_{1})$. $L_{p}^{q}(x)$ are the associated Laguerre polynomial defined as
\begin{equation}
L_{p}^{q}(x)\equiv\frac{x^{-q}e^{x}}{p!}\frac{\textrm{d}^{p}}{\textrm{d}x^{p}}\left(x^{p+q}e^{-x}\right).
\end{equation}
Two integers, $n$ and $m$ uniquely describe every member of the basis set, with $n>0$ and $|m|\leq n$. The surface brightness of a galaxy $f(\theta)$ is given by
\begin{equation}
f(\theta)=f(\theta,\phi)=\sum_{n=0}^{\infty}\sum_{m=-n}^{n}f_{n,m}P_{n,m}(\theta,\phi;\beta)
\end{equation}
The polar basis set has rotational symmetries which are very useful for describing weak lensing, so in practice galaxies are decomposed using the Cartesian basis set (which easily describes square pixels) and then transformed to the polar set. 

Both the Cartesian and polar shapelet basis functions have simple behaviour under convolution \citep{2003MNRAS.338...35R} and deconvolution \citep{2003MNRAS.338...48R}, making them particularly well suited for describing and correcting the effects of a PSF. Since this is important for our shape measurement analysis, we briefly describe deconvolution in the shapelet framework below.

\subsection{Deconvolution with Shapelets}
The approach used in this work for deconvolution is to estimate the deconvolved shapelet coefficients $f_{n_{1},n_{2}}$ by `forward convolving' the shapelet basis function with the PSF model $g(\mathbf{x})$, in advance, creating a new basis set which we label 
\begin{equation}
D_{n_{1},n_{2}}(\mathbf{x};\beta)=g(\mathbf{x};\beta)\ast B_{n_{1},n_{2}}(\mathbf{x};\beta),
\end{equation}
with an equivalent expression for the polar shapelet basis functions. Fitting the data $h(\mathbf{x})$ using this new basis set $D_{n_{1},n_{2}}$, one obtains the deconvolved shapelet model for the galaxy as follows:
\begin{eqnarray}
h(\mathbf{x})&=& g(\mathbf{x})\ast f(\mathbf{x}) \nonumber \\
&=& g(\mathbf{x})\ast\left[\sum_{n_{1},n_{2}}^{\infty}f_{n_{1},n_{2}}B_{n_{1},n_{2}}(\mathbf{x};\beta) \right]\nonumber \\
&=&\sum_{n_{1},n_{2}}^{\infty}f_{n_{1},n_{2}}\left[g(\mathbf{x})\ast B_{n_{1},n_{2}}(\mathbf{x};\beta)\right] \nonumber \\
&=&\sum_{n_{1},n_{2}}^{\infty}f_{n_{1},n_{2}}D_{n_{1},n_{2}}(\mathbf{x};\beta) 
\end{eqnarray}
Comparing with Equation \ref{eq:decomp}, the returned coefficients will reconstruct the the deconvolved image when used with the original basis set. 

We use the publicly available shapelets software package\footnote{http://www.astro.caltech.edu/$\sim$rjm/shapelets/code/} described in \citet{2005MNRAS.363..197M} in order to make shapelet decompositions for all our objects. This code is well tested using optical data (c.f. \citealp{2006MNRAS.368.1323H} and \citealp{2007MNRAS.376...13M}). In \citet{2010MNRAS.401.2572P} we indicated the applicability of this code to radio data with some modifications; with our radio image simulations, we are now in a position to demonstrate its ability to accurately recover ellipticities in \S\ref{results}. In all that follows we have used the  shapelet equivalent to Gaussian weighted quadrupole moments to estimate ellipticities for all our objects. In the polar shapelet convention this estimator takes the form 
\begin{equation}
\epsilon_{1,2}=\frac{\sqrt{2}f^{\prime}_{2,2}}{\langle f_{0,0}-f_{4,0}\rangle},
\end{equation}
as is fully discussed in \citet{2007MNRAS.380..229M}.

\subsection{Image Simulations with Shapelets}
\label{shapesims}

In this section we describe how the input images are created. We generate two different sets of input images: the first set contains point sources, consisting of a collection of single illuminated pixels at random locations. The second suite are images containing detailed galaxy shapes and morphologies, but each with a centroid at the same locations as the point sources in the previous image set. \corr{In optical weak lensing studies stars are used to study the behaviour of the PSF across the field. Since stellar point sources are not present in radio images some other mechanism will be needed to study the beam behaviour. Nonetheless, we simulate the point sources to study the behaviour of the dirty beam or point spread function of the telescope.} The galaxy images provide a more realistic challenge, with the goal being to recover the (known) shapes of galaxies in the presence of all the systematics that distort them. 
 
The galaxy images are created using the shapelets based formalism described above. The method of populating the images is based on the \citet{2004MNRAS.348..214M} pipeline, which was used as part of the STEP2 shear methods testing programme \citep{2007MNRAS.376...13M}.  The motivation for the use of shapelets galaxy models, as discussed in \citet{2004MNRAS.348..214M}, is to simulate deep sky images which include some degree of realistic galaxy morphology. 

The procedure used to generate simulated galaxy images in this paper most closely follows that described by \citet{2012arXiv1211.0966R} for the generation of simulations of \emph{Hubble Space Telescope} (\emph{HST}) data.  Using a PSF and real noise properties estimated directly from real \emph{HST} survey data (specifically the GEMS Survey: see, e.g., \citealp{2005MNRAS.361..160H}, \citealp{2004ApJS..152..163R}), \citet{2012arXiv1211.0966R} created realistic simulations of Advanced Camera for Surveys (ACS) images in the F606W filter.  These were then used to investigate shear and higher order lensing shape measurements in ACS data.

As we are simulating radio observations, we do not require a high level of similarity to optical images, in terms of noise, PSF or even shape; in our previous work, \citet{2010MNRAS.401.2572P}, we found that on a galaxy-by-galaxy basis, optical and radio shapes of galaxies are only weakly correlated. However, we also showed that  the overall distribution of ellipticities in the optical and radio sky were very similar, so we choose to follow the optical distribution of shapelets in this study. This is a reasonable choice given the current absence of a  large observed sample of highly resolved radio objects at $\mu$Jy flux density thresholds, which could inform us about the details of the radio shapelet distribution.

A `starter set' of galaxy shapelet models was created using shapelet decompositions of GEMS images, modelled according to \citet{2012arXiv1211.0966R} using a spatially and temporally varying model of the ACS PSF.  The temporal variation of each shapelet coefficient in an $n_{\rm max} = 20$ model of the PSF was modelled in three epochs as described by \citet{2005MNRAS.361..160H}, using a third order polynomial surface to describe the spatial variation on each ACS chip.  The resulting PSF-deconvolved shapelet models are then sampled from to produce the simulated galaxy images. 

When creating the simulated galaxies from the starter set, the models are randomly rotated and/or inverted to eradicate any remaining signature of gravitational lensing. We also note that the starter set represents a large but ultimately limited sample of galaxy morphologies. This is alleviated by introducing small random perturbations to the shapelet models (see \citealp{2004MNRAS.348..214M} \& \citealp{2012arXiv1211.0966R} for details).

\corr{Using this prescription, a suite of 100 images of size $1024\times 1024$ pixels with a pixel scale of $0.05^{\prime\prime}$ were created. We have constructed these images such that all the sources (galaxies and point sources) were at a constant flux. We explicitly assume that we are able to perfectly calibrate and remove the brighter sources from the data. The bright sources will be fewer in number than the far more numerous faint galaxies, which will be the most useful from a weak lensing perspective. The assumption or perfect bright source calibration could be potentially limiting for future weak leaning studies as the bright sources will dominate the side lobe contribution, however we con side assessing this contribution to the shape measurement analysis to be outside the scoot of this initial work.}

We also inspected the relationship between the $\beta$ values of the simulated galaxies, derived from GEMS, and the $\beta$ parameters that were derived from the shapelet modelling of the actual radio sources in \citet{2010MNRAS.401.2572P}. It was found that the distribution of the $\beta$ parameters from the best-fitting shapelet models of the GEMS data had a systematically lower value $(\textrm{shift of } 0.185^{\prime\prime})$ than the radio shapelet models from \citet{2010MNRAS.401.2572P} ($\langle\beta\rangle\simeq0.3^{\prime\prime}$ for the silver catalogue in that work). To make the simulations a more realistic representation of radio data we therefore add 0.185$^{\prime\prime}$ to the $\beta$ scale parameter for our galaxy models; these then provide a close match to the distribution of the derived $\beta$ values from \citet{2010MNRAS.401.2572P}.

\begin{figure}
\centering
\includegraphics[width=\linewidth]{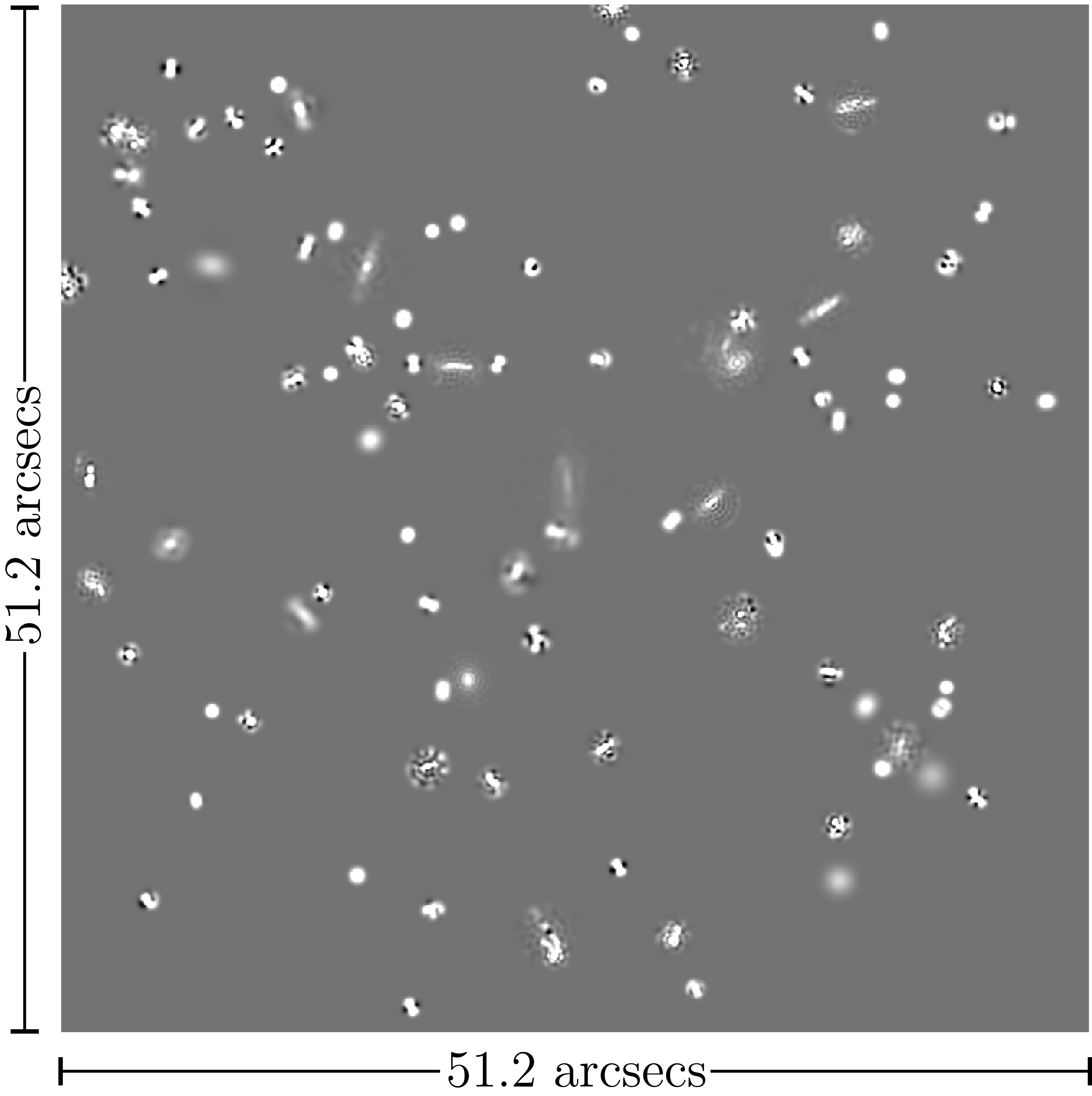}
\caption{Example of input shapelet-based galaxy images; the image has a linear greyscale.\label{fig:simeg}}
\end{figure}

For each image, we have a corresponding catalogue containing the input shapelet coefficients of each source in the image, as well as the centroid of the objects. An example of one of our images containing galaxies is shown in Figure \ref{fig:simeg}.

\corr{The negative value regions associated with some of the sources created within the images are a result of the deconvolution of the HST data. The presence of noise in the data manifests as high frequency artefacts after deconvolution is performed. If these simulated images were to be used for an optical study (e.g. \citet{2012arXiv1211.0966R}, \citet{2007MNRAS.380..229M}) then Figure \ref{fig:simeg} would be reconvolved with the HST PSF and appropriate noise added to produce a more realistic looking image.}

\corr{We note that these images represent an ambitious imaging scenario, as they are very densely populated fields representing deep pointings. The input images contain a number density of $n\sim140$ arcmin$^{-2}$, which is far larger than current lensing studies, even those in space. There is motivation for setting a high number density as it better allows us to probe systematic errors related to radio interferometers deriving from both the instrumental setup and the radio imaging pipeline. While a large number of point sources allows one to better probe the dirty beam behaviour across the field, a larger number density in the galaxies means a more challenging imaging problem, since sidelobe noise will be greater in densely packed images.}

\corr{We note that both eMERLIN and LOFAR have much larger instantaneous fields-of-view than considered here. This is due mainly to the practical computational limits we currently have. Since we need to create images with a very fine pixel resolution, to simulate large fields-of-view would be computationally challenging, the current imager is also limited in terms of the maximum size of image it can compute. Computation of the visibilities from the image also depend on the size of image so increases the length of time it takes to run the simulation. Smaller images, with many sources is favoured in terms of computation time and so this was adopted here since it is not currently possible to realistic fields-of-view at the required resolution for weak lensing.}

Note that the simulated galaxies have not been sheared; in this analysis we are not concerned with trying to recover a cosmic shear signal from galaxies; rather, we are examining how well we can recover known input ellipticities, which is the basis for recovering such a signal.

Having described the details of the input images, we now describe the details of our simulated configurations. We aim to create observations from radio arrays of our choice, including imaging steps such as dirty beam deconvolution and visibility weighting, in order to create realistic restored images. 

\subsection{Simulations Pipeline}

\begin{figure*}
\centering
\includegraphics[width=\linewidth]{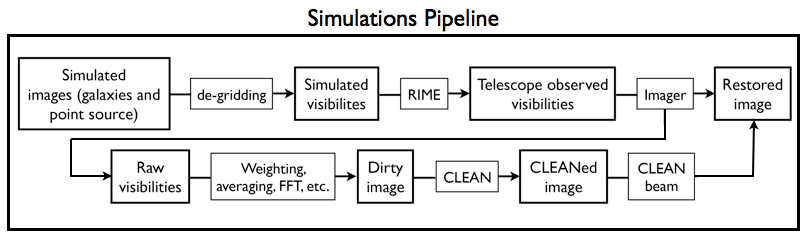}
\caption{Summary of the main steps involved in the simulation pipeline from start to finish.\label{fig:simpipeline}}
\end{figure*}

Our simulation pipeline is implemented using the MeqTrees\footnote{\href{http://www.astron.nl/meqwiki}{http://www.astron.nl/meqwiki}} software system. MeqTrees is a software package for implementing the Measurement Equation for radio interferometers and is fully described in \citet{2010A&A...524A..61N} and \citet{2011A&A...527A.106S}. The heart of a Measurement Equation is formed by the $2\times 2$ Jones matrices \citep{JONES:41} which describes the various effects associated with observations that can corrupt the measured visibilities. The formulation for a generic Radio Interferometer Measurement Equation (RIME) was developed by \citet{1996A&AS..117..137H}, after preparatory work by \citet{1964ApJ...139..551M}. \citet{2000A&AS..143..515H} then recast the formalism into $2\times 2$ matrix form which is used within MeqTrees\footnote{Some versions of the RIME are still implemented using the $4\times4$  Mueller matrices \citep{1948JOptSocAm38.661}, which are entirely equivalent.}. The RIME provides an elegant mathematical framework for generic radio instruments, both existing and future, to be better understood, simulated and calibrated. For a full description of the mathematical formalism for the measurement equation of interferometers we refer the reader to \citet{1996A&AS..117..137H}, \citet{2011A&A...527A.106S} and references therein. 

The MeqTrees software was originally designed to implement the measurement equations for the purposes of simulation and calibrating \citep{2010A&A...524A..61N}. We use MeqTrees in this work to simulate the sky as would be observed by a specific radio interferometer.The galaxy images created above are fed into the MeqTrees simulator, which calculates observed visibilities; these visibilities are then imaged. The user can specify many options such as the level of noise on the visibilities, the specific method used to deconvolve the dirty beam, and the weighting scheme applied to the visibility data prior to imaging. 

One subtlety is that the input in our work is an image rather than a set of visibilities. In order to obtain $uv$ samples from the image, we use a module of the MeqTrees software that implements a `de-gridding' algorithm. This algorithm is an interpolation scheme that allows one to transform between the regularly gridded image and a sparsely sampled Fourier (or $uv$) plane. The details of how this algorithm work is explained fully in \citep{SZE....1986....GRIDDING}. 

\corr{The well known general relationship between observed visibilities and the true sky brightness for an interferometer is given by: 
\begin{equation}
\label{eq:simvis}
I(\ell,m,n)=\int\int V(u,v,w)\frac{e^{2i\pi(u\ell+vm+w(n-1))}}{\sqrt{1-\ell^{2}-m^{2}}}\textrm{d}u\textrm{d}v,
\end{equation}
where $n=\sqrt{1-\ell^{2}-m^{2}}$. This expression can be reduced to a two dimension Fourier Transform if the field-of view is small \citep{1999ASPC..180....1C}, i.e. if $n\simeq1$. Since our study consists of fields that are less than an arc minute on a side we have adopted this simplification here such that: 
\begin{equation}
\label{eq:simvis}
I(\ell,m)\simeq\int\int V(u,v)e^{2i\pi(u\ell+vm)}\textrm{d}u\textrm{d}v,
\end{equation}}Equation \ref{eq:simvis} should also include the {\it sampling function} term, $S(u,v)$ on the left hand side. The sampling function accounts for the fact that the visibilities are collected at a set of discrete locations in the $uv$ plane. Its functional form is simply a linear combination of Dirac delta functions at all $(u,v)$ where data is collected. The Fourier inversion of Equation \ref{eq:simvis} leads to the {\it dirty image}, which is the convolution between the sampling function (in the image plane) and true sky brightness. 
Since Direct Fourier Transforms become computationally expensive with large datasets, Fast Fourier Transforms (FFTs) need to be used. Although the use of FFTs speeds up computations, it requires that we grid the data. The process by which this de-gridding is done, and its implication for the recovered image is described in \citet{SZE....1986....GRIDDING}. The MeqTrees software employs the use of prolate spheroidal functions in the de-gridding process, which have been shown to reduce the effect of aliasing \citep{SZE....1986....GRIDDING}. 

In Figure \ref{fig:simpipeline} we illustrate the simulation setup from start to finish. (Note that in this work we have focused on one particular method of deconvolution of the dirty beam, namely the CLEAN algorithm \citep{1974A&AS...15..417H}, which we shall describe in Section \ref{sec:clean}). We start with an image that is Fourier transformed to create the set of simulated visibilities. These are then modified by the RIME formalism for our given set of parameters to create a set of visibilities as observed by our given array. These visibilities are Fourier transformed again to produce the dirty image, which is then CLEANed as our setup defines to produce the CLEAN component image. The residual of the dirty image is then added to the CLEAN components once convolved with the CLEAN beam (usually an elliptical Gaussian measured from the main lobe of the dirty beam). The final step involves convolving the CLEANed image with the CLEAN beam. The CLEAN beam is usually a Gaussian fitted to the main lobe of the PSF and it is this final beam that the point source simulations will characterise. This is also what we shall deconvolve from the galaxies to produce our final images from which the shapes of our sources will be measured.   

 \subsection{Simulated Configurations}
 \label{sec:config}

\begin{figure*}
\centering
\includegraphics[width=\linewidth, height=1.1\linewidth]{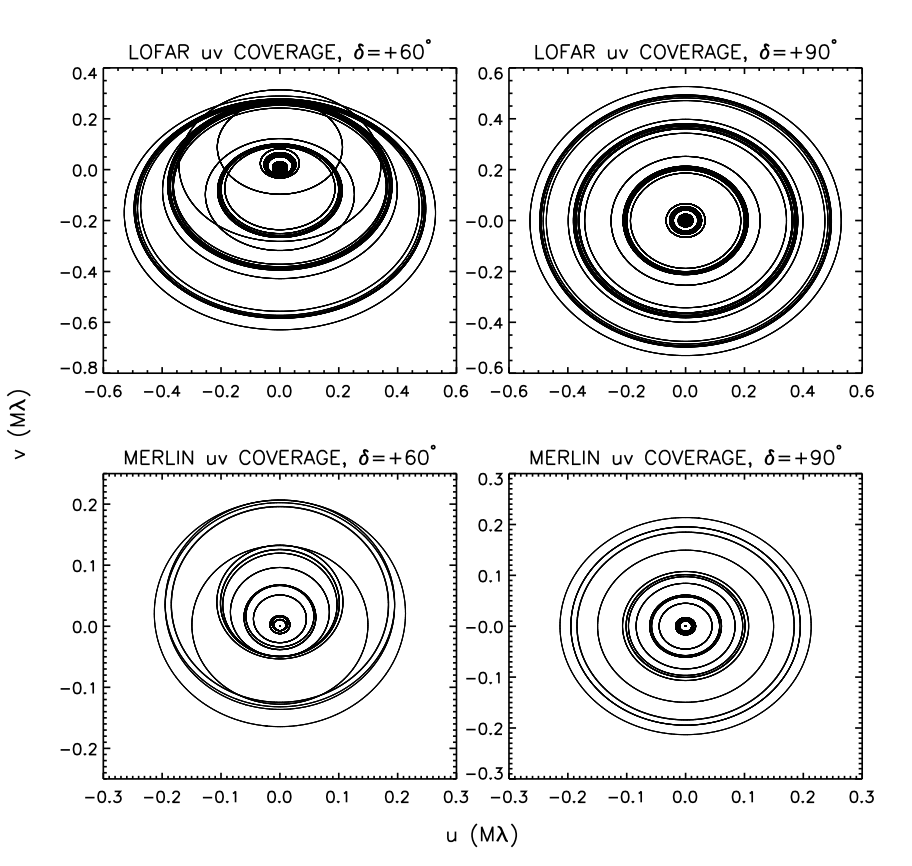}
\caption{$uv$ coverage corresponding to our 4 starting MSs. The top panels show LOFAR coverage for observations at two declinations, while the bottom panels show eMERLIN coverage. \label{fig:uvcoverage}}
\end{figure*}
 
In this section we describe the different experiments we have carried out to examine a variety of potential causes of systematic effects. 

We choose to concentrate on two specific interferometers: eMERLIN and LOFAR. The eMERLIN array is an upgrade to the existing MERLIN array, maintaining the same number of dishes. The eMERLIN array upgrade is designed to increase sensitivity by more than an order of magnitude by using new receivers and telescope electronics. The array spans 217 km, with observing bands at 1.3-1.8 GHz, 4-8 GHz and 22-24 GHz, with a total bandwidth of 4 GHz. It also has resolution capabilities of between 10-150 mas and sensitivity of $\sim1\mu$Jy. 

LOFAR is a new generation radio interferometer, with the ultimate goal of surveying the Universe at low frequencies with high resolution and sensitivity. It operates in the less explored low frequency range of 10-250 MHz. LOFAR belongs to a new generation of telescopes, with the concept of utilising many inexpensive dipole antennae arranged in stations without any moving parts, in contrast to the usual notion of radio dishes as used by eMERLIN. Different parts of the sky are observed by steering the beam electronically. \corr{The spatial resolution of the Netherlands part of the array is governed by the $\sim100$ km baselines, leading to a resolution of $\sim2$ arcseconds at 240 MHz. The LOFAR array will eventually be extended over Europe, with stations in the UK, Germany, Sweden and France. These Europe wide baselines will reach $\sim1500$km, leading to  $\sim0.14$ arcsecond resolution at 240 MHz. In our study we include some of these larger $(< 700$ km), Europe-wide baselines.} 

For both arrays we create two different Measurement Sets\footnote{Interested readers are referred to \href{http://aips2.nrao.edu/docs/notes/229/229.html}{http://aips2.nrao.edu/docs/notes/229/229.html} for a detailed explanation.} (MSs) centred on a fixed RA $00^{\mbox{h}}02^{\mbox{m}}34^{\mbox{s}}.43$, but with differing DEC $+90^{\circ}16^{\prime}41^{\prime\prime}.75$ and DEC $+60^{\circ}16^{\prime}41^{\prime\prime}.75$. A Measurement Set is a specific definition of how visibility data is stored, designed to be compatible with the measurement equation formalism. For our purposes we can think of it as a large table containing information about the particular observation in question. For the eMERLIN MSs we use a central frequency $\nu=1.4$ GHz while for the corresponding LOFAR MSs we use a central frequency $\nu=240$ MHz. In both cases we set our bandwidth specifications to 128 channels each of 125 kHz resulting in a 16 MHz bandwidth. We have also employed a 20 sec time averaging over a 24 hour observation. \corr{We further note here that we have assumed that the time and frequency sampling is fine enough to keep smearing effects to a minimum and are not considered in this study. In practice, these smearing effects will place some practical limit of the field-of-view that can used for the final analysis. In assuming that the smearing effects are small we are effectively assuming that the majority of the field-of-view is available for the analysis.} The $uv$ coverage corresponding to these four MSs are shown in Figure \ref{fig:uvcoverage}. 

We use the MeqTrees imager to create restored (deconvolved) images for these telescope configurations. We image an area twice the size of the original input image, and use a portion of this extra image to estimate the rms of the residuals as described below. Our default imaging option uses the Clark CLEAN deconvolution algorithm; we shall describe below how we estimate the number of CLEAN components used for each simulation. In creating the restored images we have kept our pixel scale the same as that of the original input images, $0.05^{\prime\prime}$.

\corr{For all of the simulations that we perform we create a set of 10 MSs and then use each to simulate and image 10 point source and 10 galaxy images each. The eMERLIN and LOFAR MSs described above are 1.5 GB and 17 GB respectively. Each eMERLIN MS took ~ 36 hours to simulate and image its 20 images while for the LOFAR case this was closer to 60 hours. }

In some runs of the analysis we do not include measurement noise (i.e. the visibilities are not corrupted in any way). Clearly this is unrealistic, but allows us to distinguish between systematic effects and noise effects. Once we have analysed these noise-free images, we explore the effect of adding measurement noise to see what effect that has on our results. Below we list and briefly discuss the telescope effects we study in this work. 

\subsubsection{Observing Angle}
\label{sec:obsang}
We have created MSs for both arrays observing at differing declination angles to examine what effect this has on our shape measurements. It can be shown that the equation for the ellipse traced in the $uv$ plane by a particular baseline is given by \citep{1999ASPC..180...11T}
\begin{equation}
u^{2}+\left( \frac{v-(L_{Z}/\lambda)\cos \delta_{0}}{\sin \delta_{0}} \right)^{2}=\frac{L_{X}^{2}+L_{Y}^{2}}{\lambda^{2}},
\end{equation}
where $L_{X,Y,Z}$ are the coordinate separations between the two antennae and $\delta_{0}$ is the declination of the phase tracking centre (usually where the field-of-view is centred). As the interferometer observes a point on the celestial sphere, the rotation of the Earth causes the $u$ and $v$ components of the baseline to trace out an elliptical locus according to this equation. 

Since the sampling function is effectively a collection of these elliptical loci (c.f. Figure \ref{fig:uvcoverage}), we see that it depends on the declination of the observation and the antenna spacings.  The sampling function, as already mentioned, determines the PSF (or dirty beam) for the experiment. Our different declination observations give us the means to see how the variation of the sampling function (measured from the point sources) affects the recovered galaxy shape distributions. 

\subsubsection{CLEAN}
\label{sec:clean}
We have mentioned above the Fourier relationship between the observed visibilities and the desired sky intensity. We discussed how the FFT of the visibilities leads to an estimate of the dirty image, which is the sky brightness convolved with the sampling function. In order to obtain an estimate of the sky intensity we need some way to perform a deconvolution. 

The commonly used algorithm CLEAN is one  way to perform this deconvolution. Devised by \citet{1974A&AS...15..417H} it assumes that the radio sky can be represented by a collection of point sources (CLEAN components) in an otherwise empty field. The intensity distribution $I(\ell,m)$ is then approximated by superposition of these point sources which have a positive intensity $A_{i}$, at the locations $(\ell_{i},m_{i})$. The CLEAN algorithm then aims to determine $A_{i}(\ell_{i},m_{i})$ such that
\begin{equation}
I^{D}(\ell,m)=\sum_{i}A_{i}B(\ell-\ell_{i},m-m_{i})+I_{\epsilon}(\ell,m),
\end{equation}
where $I^{D}(\ell,m)$ is the dirty image that is obtained from the inversion of the visibilities, $B(\ell,m)$ is the dirty beam which is the inverted sampling function and, $I_{\epsilon}(\ell,m)$ is the residual brightness distribution. The approximation is deemed to have been successful if this residual noise is similar to that of the measured visibilities. This decomposition cannot be analytically computed and an iterative approach is required. 

The original CLEAN algorithm is applied entirely in the image plane. It uses a simple iterative procedure to find the position and strengths of the sources in the dirty image, from which a dirty beam multiplied by the peak strength is subtracted. All positions where this occurs are recorded as well as the corresponding peak flux. The procedure stops when all remaining peaks fall below some specified level. The recorded positions and fluxes constitute the point source model, which is then convolved with the idealised CLEAN beam (usually the central lobe of the dirty beam) and the residuals from the dirty image are added to produce the final, deconvolved image. We refer the reader to \citet{1978A&A....65..345S} and \citet{1979ASSL...76..261S} for further details.

Our imaging pipeline can implement a variety of CLEAN algorithms; in this paper we choose a widely used version, the Clark CLEAN \citep{1980A&A....89..377C}, which efficiently implements the algorithm and can provide improved speed for larger images. 

\corr{We note that our imaging pipeline makes use of the light weight imager (lwimager) based on the CASA imaging libraries. Throughout the analysis of the simulations in this work we have adopted a cleaning threshold of 0 and a loop gain parameter of 0.1.}

It is important for us to study the effect of the CLEAN method (and the process of deconvolution as a whole) on shape measurement. Although CLEAN is well tested and appropriate for many imaging applications, it is non-linear and does not necessarily converge in a well-defined manner. It is not immediately clear if it is suitable for the purpose of shape measurement as required for weak lensing; we will test this in the next section.

\section{Shape Measurement Results}
\label{results}
In this section we present the results from the shape measurement analyses of both point sources and galaxies. We firstly address the question of how many CLEAN components are required to adequately represent these images. We then examine the behaviour of point source ellipticities, for different telescope configurations and declinations. Next we assess the modelling of galaxy shapes, in relation to the shapelet scale parameter and uv sampling. All this is in preparation for the main result of the paper, which is the presentation of whether output ellipticity estimates are in line with true input ellipticities. Finally, we discuss the impact of realistic noise on the measurement of ellipticity.

\subsection{Required Number of CLEAN Components}

\begin{figure*}
\centering
\includegraphics[scale=0.75]{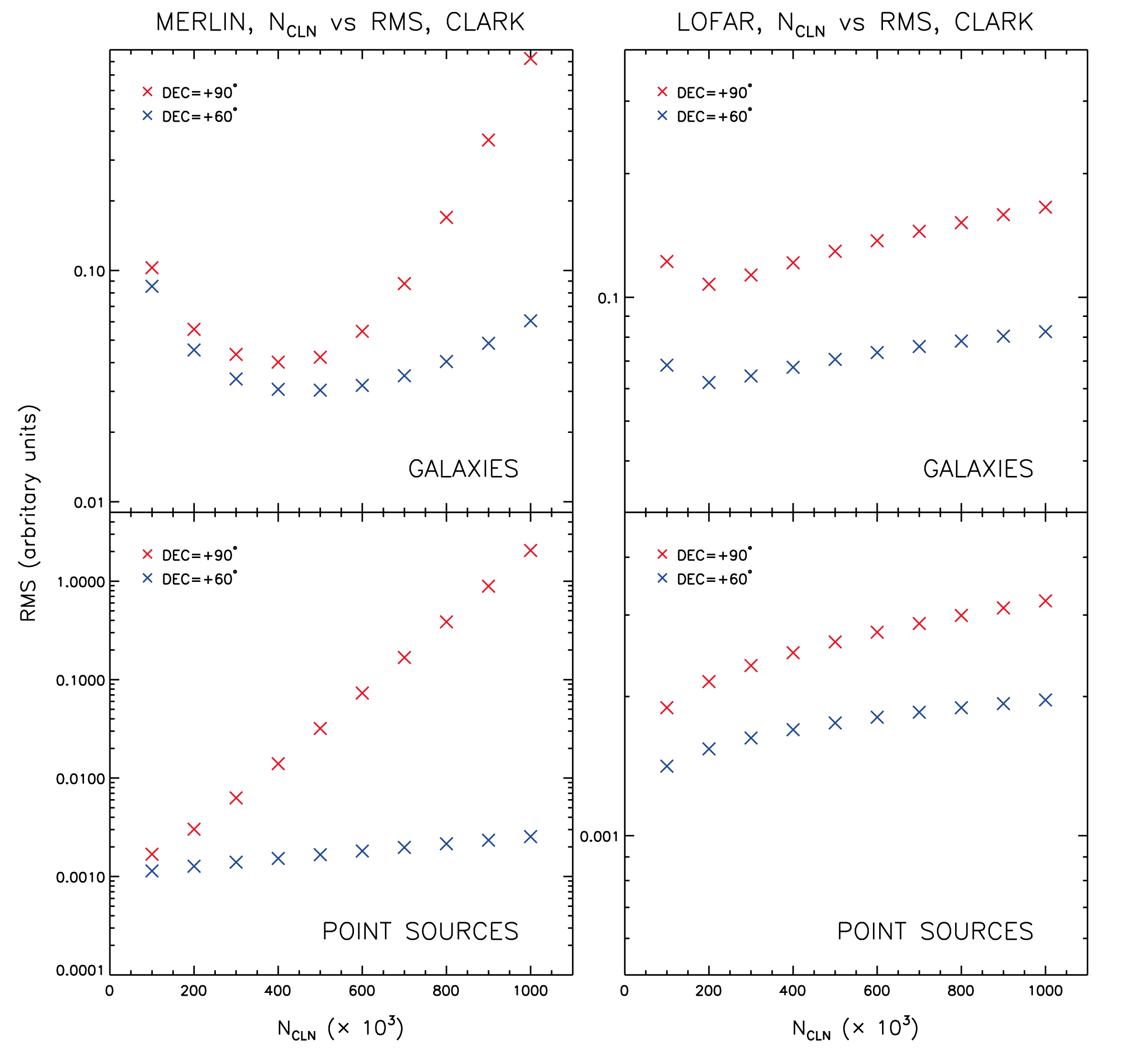}
\caption{Mean rms residuals for varying number of Clark CLEAN components, for point sources and galaxies, in each of our four MSs. \label{fig:clean}}
\end{figure*}

We aim to find the lowest number of components that suitably deconvolve the sources in the images, since this reduces computation time and avoids fitting noise. To assess this issue, we set several different target numbers of CLEAN components on one of the images from our set of simulations. We ran this experiment on all four MSs with the point sources and the galaxies. Since the point sources have a much simpler set of source structures, it would be expected that they would require fewer CLEAN iterations. We examine the residuals (of the dirty image) in order to compare the relative merits of the different numbers of CLEAN components used. We examine where we find the minimum rms residuals as a function of number of CLEAN iterations.

We image the residuals just outside the patch where we have sources in order to estimate the residual side lobe noise from the sources inside the central part of the image. To estimate the rms of the residuals, we select a frame around the inner image containing the sources, with a width 16 pixels $(0.8^{\prime\prime})$ which we split into 20 cells, from which we calculate a mean rms. In Figure \ref{fig:clean} we show how the mean point source and galaxy rms residuals vary with the number of CLEAN components for each of our MSs.

We wish to perform our analyses on the images with the smallest rms residuals. For  the point source images, the lowest tested number of CLEAN components, $N_{CLN}=10^{5}$, has the smallest rms, and we adopt this value. We see that in the eMERLIN galaxy simulations there is a different minimum in the rms residuals for the two different declinations with $N_{CLN}=4\times10^{5}$ and $N_{CLN}=5\times10^{5}$ favoured by the DEC$=+90^{\circ}$ and DEC$=+60^{\circ}$ simulations respectively. For the LOFAR cases the minimum lies near $N_{CLN}=2\times10^{5}$; these are the iteration numbers that we have adopted for galaxies.

\subsection{Source Extraction and Shapelet Modelling}
Since the images we create have known source positions, we bypass the source extraction stage of a lensing analysis. We instead use SExtractor in the ASSOC mode whereby we input a catalogue of positions where objects are known to lie. Of course, even in this mode we will only detect objects which meet the internal requirements to be classified as an object. We use the default SExtractor settings of Table \ref{tab:sexsettings}; with these settings we find that we are able to detect $\simeq85-90\%$ of the objects that are in the original images, with detected number densities ranging from $n\simeq120-130$ arcmin$^{-2}$.

\begin{table}
\begin{minipage}{90mm}
\centering \caption{Table summarising main SExtractor parameters. \label{tab:sexsettings}}
\begin{tabular}{cc}\hline
Configuration Parameter & Value  \\ \hline \hline
$\textrm{ASSOC\_TYPE}$ & NEAREST \\
$\textrm{ASSOC\_RADIUS}$ & 5 pixels \\
$\textrm{DETECT\_THRESH}$ & 1.5 \\
$\textrm{DETECT\_MINAREA}$ & 5 \\
$\textrm{DEBLEND\_MINCONT}$ & 0.005 \\
$\textrm{DEBLEND\_NTHRESH}$ & 32 \\
$\textrm{BACK\_SIZE}$ & 64 \\
$\textrm{BACK\_FILTERSIZE}$ & 3 \\
\hline\hline
\end{tabular}
\end{minipage}
\end{table}

We use the shapelets software as described in \S\ref{shapelets} to estimate the ellipticities of the point sources and the galaxies. We first examine the point sources, in order to estimate the beam, which we can then use to deconvolve from the galaxies. After this, the deconvolved galaxy ellipticities will be compared to the ellipticities of the original catalogues.  

\subsection{Point Sources}
\label{sec:point}
\begin{figure*}
\centering
\includegraphics[scale=0.4]{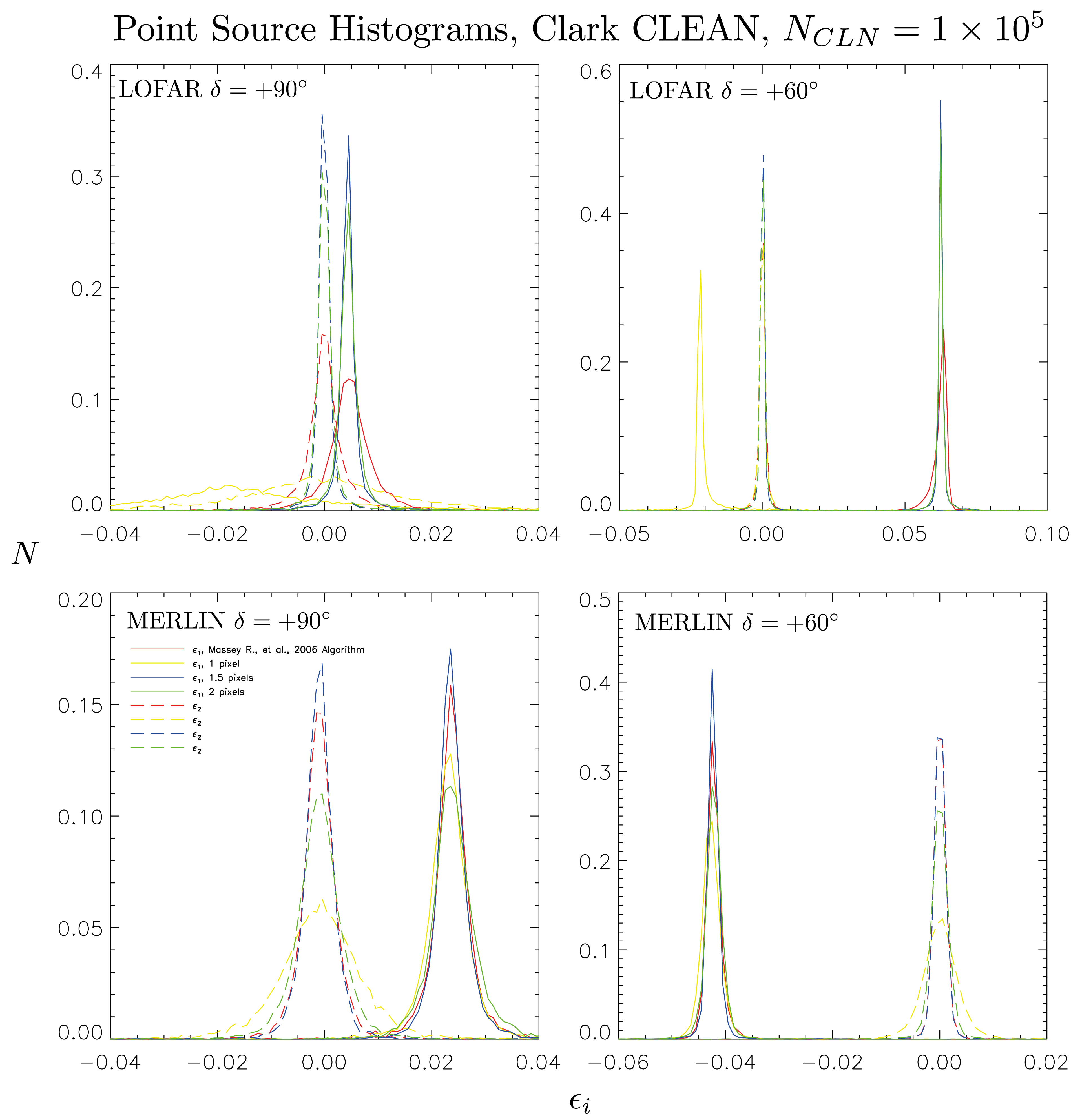}
\caption{Normalised ellipticity distributions for our four default measurement sets, for different methods of shapelet decompositions. Red curves correspond to the shapelet optimisation algorithm of \citep{2005MNRAS.363..197M}, and the yellow, blue and green curves correspond to decompositions done with fixed $n_{max}=10$ and $\beta=1, 1.5, 2$ pixels respectively. The solid lines show ellipticity element $\epsilon_{1}$ and the dashed show $\epsilon_{2}$. \label{fig:delhists}}
\end{figure*}

For the point sources we use a straightforward approach for shapelet modelling, decomposing each source into shapelets while examining how the distribution of ellipticities changes if we vary the $\beta$ parameter. We try the algorithm employed in the shapelet software \citep{2005MNRAS.363..197M} to optimise the $\beta$ and $n_{max}$ parameters, as well as adopting $\beta$ values of 1, 1.5 and 2 pixels with a fixed $n_{max}=10$. We find the mean point source FWHM is approximately $\sim2$ pixels for the eMERLIN simulations and $\sim6$ pixels for the LOFAR ones (where the pixel size is chosen to be the same as for the input images). In Figure \ref{fig:delhists} we show the normalised ellipticity distributions of the point sources for each measurement set; note that the `Massey' label refers to ellipticities obtained using the optimisation algorithm in the shapelets software \citep{2005MNRAS.363..197M}.

From these ellipticity distributions we can see that the most tightly peaked distributions are produced from a fixed $\beta=1.5$ pixel approach. The underlying PSF ellipticity distribution should be narrow in a small image region, so we adopt $\beta=1.5$ for our PSF measurements in all four cases. We also note that in the LOFAR ellipticity distributions, changing $\beta$ can change the measured ellipticity of the PSF and therefore would lead to different galaxy ellipticities if used for deconvolution. This can be understood by comparing the mean point source FWHM of 6 pixels with the attempted $\beta$ choices; having a small $\beta$ will mean that the decomposition cannot capture the larger scale shape information, so the $\beta=1$ pixel histograms behave erratically. Since the eMERLIN mean point source FWHM is $\sim2$ pixels, all our attempted $\beta$ choices produce well measured point source models. 

\begin{figure*}
\centering
\includegraphics[width=1.3\linewidth,angle=90]{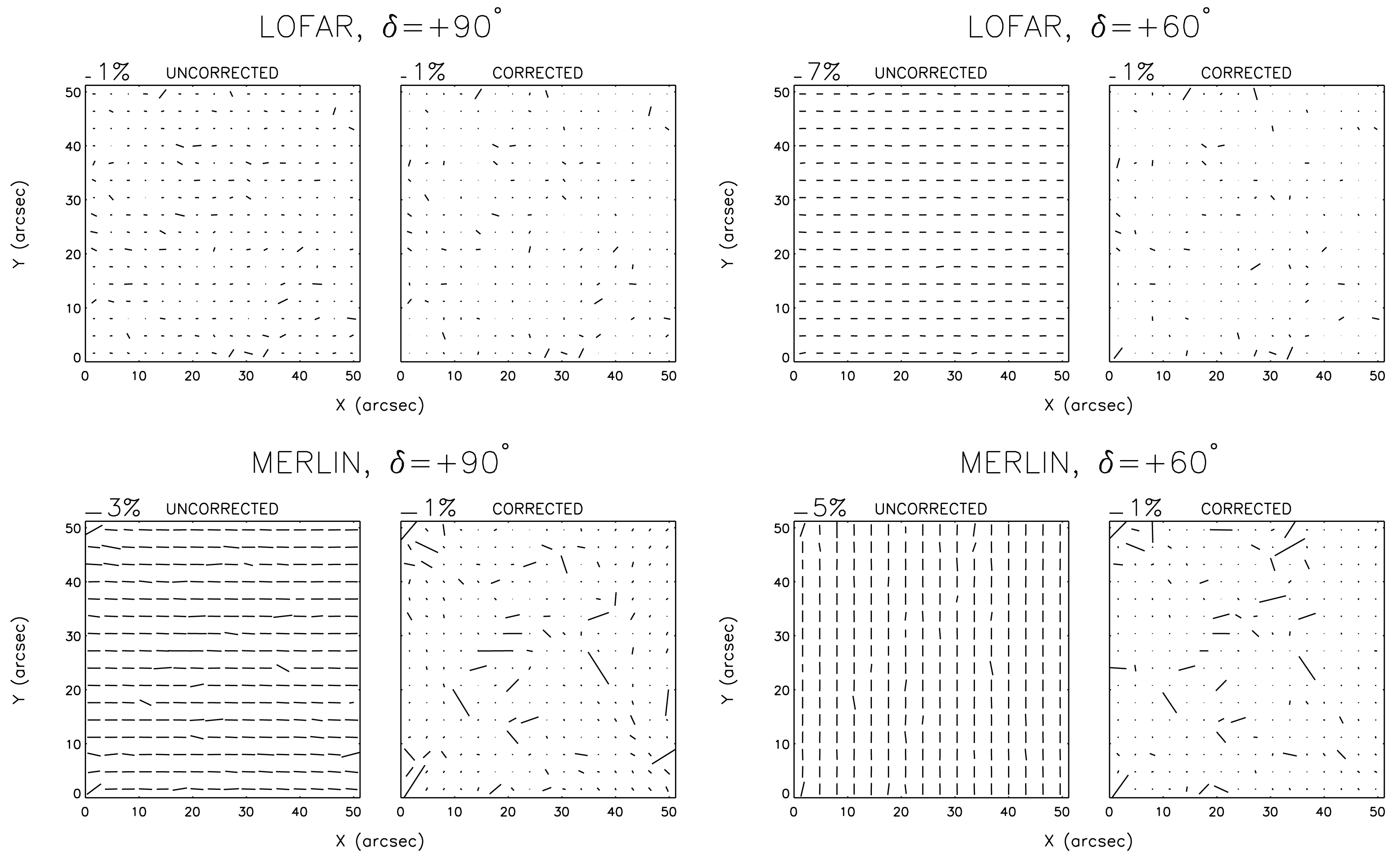}
\caption{Whisker plots of point source ellipticities across the field for our four measurement sets, for both uncorrected (lower panels in each MS) and corrected ellipticities (upper panels in each MS).\label{fig:whiskers}}
\end{figure*}

To quantify any spatial variation in PSF ellipticity, we combine all the point source simulations and bin objects'  ellipticities in $x$ and $y$ coordinates. In Figure \ref{fig:whiskers} we show how the mean ellipticities vary as a function of pixel position across our field. For each of our four MSs we show the measured (uncorrected)  and  corrected ellipticities;  the correction applied in each case is a simple subtraction of the global mean measured ellipticity on the image. 

\corr{As mentioned in Section \S\ref{sec:obsang} the dirty beam is related to the $uv$ plane sampling function. In Figure \ref{fig:whiskers} we see how the different $uv$ tracks related to our 4 MSs (as shown in Figure \ref{fig:uvcoverage}) create different real space ellipticities.}  

In all four cases, we see that there is no substantial variation in the beam behaviour across the images. In each case we find that $\epsilon_{2}\simeq0$. For the LOFAR DEC $=+90^{\circ}$ there is a small $\epsilon_{1}$ component with mean 0.005, while in the lower declination LOFAR simulation we see there is an almost constant $\epsilon_{1}=0.06$ across the field. After correction, there is no evidence of a coherent ellipticity pattern. For the eMERLIN observations, we see a change in sign of the magnitude of $\epsilon_{1}$ from positive to negative for the DEC $=+90^{\circ}$ and DEC $=+60^{\circ}$ cases respectively. As in the LOFAR case, the induced ellipticities are constant in our small fields, so a simple correction by the mean works well. 

\corr{We also note significant PSF ellipticity ($\epsilon_1=0.024$) for the
eMERLIN $\delta=90\degr$ case, which is surprising given the nominally
circular $uv$-coverage. We have found that this is induced by uniform
weighting; a naturally-weighted image yields a perfectly circular PSF
as expected. A possible explanation is given by the way uniform
weighting is traditionally implemented: the $uv$-plane is divided into
rectangular cells, and weights are assigned per cell according its
population, i.e. the number of ungridded visibilities falling into
that cell. The $uv$-cell size is determined by the size of the output
image, and the number of $uv$-cells by the resolution. This means that
the $uv$-tracks are weighted in a non-smooth manner given by the cell
boundaries, and it is possible to induce ellipticity in the PSF where
there was none in the unweighted $uv$-coverage. Sparser $uv$-coverages
should be more susceptible to this effect, which explains why the
LOFAR $\delta=90\degr$ case shows a far smaller induced ellipticity.
To confirm this, we have also experimented with different image sizes
and resolutions, and found that image size does significantly change
the measured PSF ellipticity, while resolution has little to no
effect. Our conclusion is that uniform (and robust, since it uses the
same principles) weighting will always induce a PSF ellipticity,
especially with sparser arrays. For weak lensing measurements, we can
treat this as a constant bias. In principle, more advanced weighting
methods such as that suggested by \citet{2013ExA....36...77B} should eliminate this
effect, but no mainstream imaging software currently implements them.}

We can now use these models from the point sources to construct a PSF in each of our four measurement sets and use them to deconvolve the galaxies. The peaks in the histograms provide us with a tight measure of  the beam ellipticity, as does using the mean of the uncorrected whisker plots; the measured ellipticities for the beam are summarised in Table \ref{tab:dels} for our four MSs. To construct our PSF models, we stack all our point sources and perform a shapelet decomposition with $n_{max}=10$ and $\beta=1.5$ pixels on the stacked source. 

\begin{table}
\begin{minipage}{90mm}
\centering \caption{Table summarising PSF ellipticities. \label{tab:dels}}
\begin{tabular}{ccccccccc}\hline
MS & \multicolumn{2}{c}{PSF Ellipticities} \\ & $\epsilon_{1}$ & $\epsilon_{2}$  \\ \hline \hline

LOFAR $\delta=+90^{\circ}$ & 0.005 & $-10^{-4}$ \\
LOFAR $\delta=+60^{\circ}$ & 0.063 & $-10^{-6}$  \\
eMERLIN $\delta=+90^{\circ}$ & 0.024 & $-0.001$ \\
eMERLIN $\delta=+60^{\circ}$ & -0.042 & $-10^{-5}$ \\
\hline\hline

\end{tabular}
\end{minipage}
\end{table}

\subsection{Shapelet Modelling the Galaxies}
\label{sec:gals}
In \citet{2010MNRAS.401.2572P} we found that the optimisation algorithm from \citet{2005MNRAS.363..197M} performed poorly on galaxies in our radio data, and that fixing the $n_{max}$ and $\beta$ parameters in the shapelet modelling gave much better reconstructions. We attributed the poor performance of this algorithm to the properties of the noise in the radio data. Here, we have further tested this by performing the shapelet decompositions of galaxies in a number of different ways: we have used the optimisation algorithm from \citet{2005MNRAS.363..197M} to see if it performs better on the noiseless simulated images; we have also performed several shapelet decompositions with a fixed $n_{max}=10$ and varying $\beta$; we have used multiples of the SExtractor FWHM for this purpose, analogously to \citet{2010MNRAS.401.2572P}.

As with the point sources, we use SExtractor in ASSOC mode for  source extraction before performing the shapelet decompositions. Before comparing to the original input ellipticities we removed all shapelet model failures. Most of these shapelet models can be attributed to objects being close to other objects and also objects lying near the edge of the images. Due to the high number density of objects in the images there is a large failure rate; in all four cases we lose $\simeq50\%$ of objects due to bad shapelet models. 

For all our catalogues we compute the deconvolved ellipticity estimator described in \S\ref{shapelets}, with a PSF kernel with $\beta=1.5$ pixels. We then match our catalogues to the original catalogues and compute the corresponding `true' ellipticities, $\epsilon_{i}^{t}$ to which we will compare. We have binned the catalogues in $\epsilon_{i}^{t}$ and have calculated the median measured $\epsilon_{i}$ in each bin, along with associated error. 

\begin{figure*}
\centering
\includegraphics[scale=0.375]{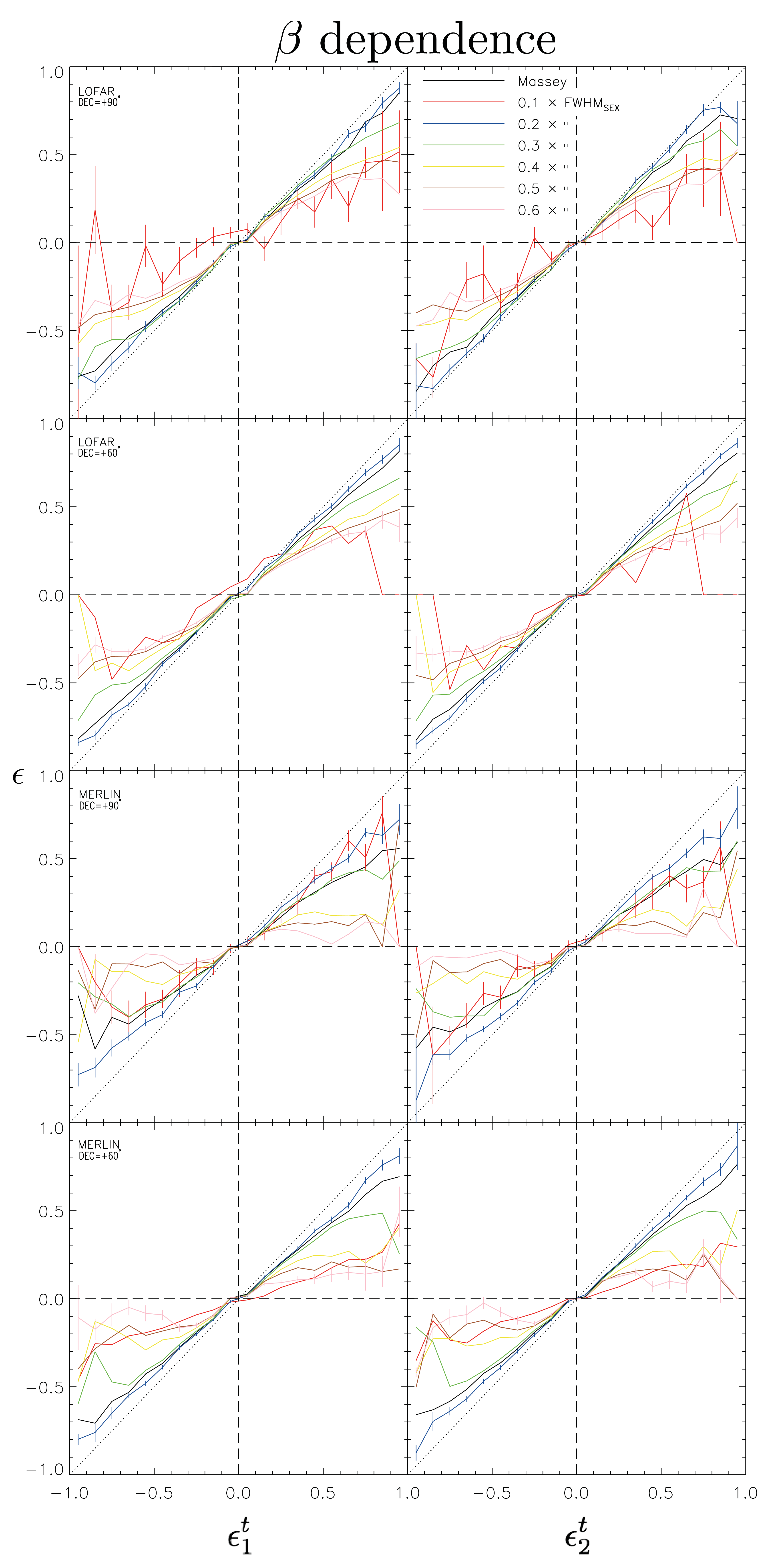}
\caption{Galaxy ellipticity, comparing input and measured ellipticities, for different choices of $\beta$  for our four configurations. Black curves correspond to the shapelet optimisation algorithm of \citep{2005MNRAS.363..197M}, and the red, blue, green, yellow, brown and pink curves correspond to decompositions done with fixed $n_{max}=10$ and $\beta=0.1, 0.2, 0.3, 0.4, 0.5 \textrm{ and } 0.6 \times \textrm{SExtractor FWHM}$ respectively. \label{fig:betaplots}}
\end{figure*}

We can now assess the best choice for the $\beta$ parameter for galaxy shapelet decompositions. For each choice of $\beta$ we fit a linear model to our data points, $(\epsilon_{i}-\epsilon_{i}^{t}=m_{i}\epsilon_{i}^{t}+c_{i})$ and compare the relative merit of each choice through the calculated $m_{i}$ and $c_{i}$ parameters. In Figure \ref{fig:betaplots} we show how the measured ellipticites, after deconvolution, compare to the input ones for different choices of fixing $\beta$. For clarity we have suppressed the error bars on all the curves barring those with the best and worst fitting models.

In all four cases we see that there is a strong dependence on the choice of $\beta$. We see that in the eMERLIN simulations, the \citet{2005MNRAS.363..197M} optimisation algorithm performs badly in comparison to fixing $n_{max}$ and $\beta$. In the LOFAR cases it performs a little better, but fixing the parameters in question is still a marginal ($\simeq 5\% \textrm{ better on } m_{i}$) improvement. The choice of $\beta$ favoured in this approach is $0.2\times \textrm{SExtractor FWHM}$; smaller $\beta$ does not capture the information about shape at the edges of objects, while larger $\beta$ smears out the detail in the object. We adopt  $\beta=0.2 \times \textrm{FWHM}$ for the rest of this work.   

\subsection{Recovered Ellipticity Distributions}

We now turn to the main result of this work, which is to demonstrate how well we can recover the input ellipticities given the data reduction and imaging pipelines described above. We have described how we obtain measurements of input ellipticity $\epsilon_{i}^{t}$ and measured ellipticities $\epsilon_{i}$; now we compare the two.

\begin{figure*}
\centering
\includegraphics[scale=0.5]{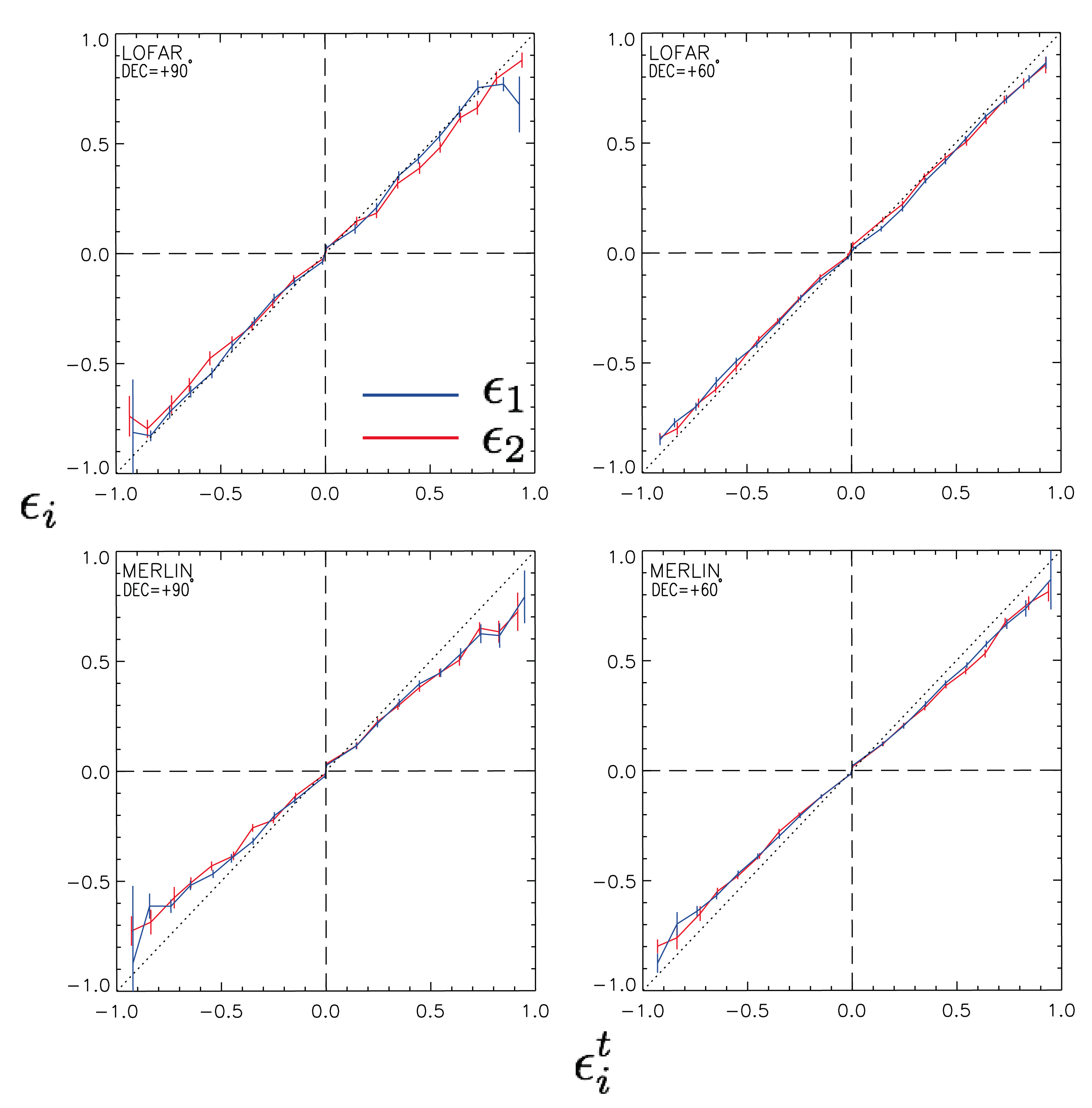}
\caption{Comparison of original and recovered ellipticities for our four configurations. \label{fig:eineout}}
\end{figure*}

In Figure \ref{fig:eineout} we show how our four MSs perform in this comparison (these are the blue curves in Figure \ref{fig:betaplots}). For each of the four cases, we compute the Pearson's correlation coefficient to quantify the correlation between the input and measured ellipticities, and also calculate the best fitting linear model $(\epsilon_{i}-\epsilon_{i}^{t}=m_{i}\epsilon_{i}^{t}+c_{i})$ as well as a $\chi_{dof}^{2}$ to assess the goodness of fit, all of which we summarise in Table \ref{tab:gals}.

$m_{i}\neq 0$ is indicative of a calibration bias, which usually results from poor correction of factors that circularise images, such as poor PSF correction. A non-zero value for $c_{i}$ suggests a systematic that induces some constant ellipticity so that even circular objects appear to have some ellipticity. 

Encouragingly we find that in all four of our test cases, we are able to recover tightly correlated input-to-output ellipticities, with the Pearson correlation coefficient being close to one in all cases (see Table \ref{tab:gals}). From Figure \ref{fig:eineout} we see that for both LOFAR and eMERLIN we find a very close relationship between the original and measured ellipticities. The fact that there is essentially no $c_{i}$ component to any of the simulations tells us that there is no constant induced ellipticity in our pipelines. There is evidence of a small calibration bias (i.e. non-zero $m_i$) in all four sets of simulations, with the LOFAR ones faring slightly better than the eMERLIN counterparts. 

We see that in both cases, the lower declination simulations provide a better recovered ellipticity (comparing $m_{i}$ values) over the DEC $=+90^{\circ}$ observations. The most accurately recovered ellipticity distribution is for  LOFAR DEC $=+60^{\circ}$, which we can see from Figure \ref{fig:uvcoverage} has the fullest $uv$ coverage, which should naturally lead to better quality images.
The worst performing simulation is eMERLIN DEC $=+90^{\circ}$, which has the simplest $uv$ coverage (c.f. Figure \ref{fig:uvcoverage}).

\begin{table*}
\begin{minipage}{180mm}
\centering \caption{Table summarising ellipticity measurement fidelity, for all the reported telescope configurations.\label{tab:gals}}
\begin{tabular}{ccccccccc}\hline
MS & DEC $(\delta)$ & $N_{CLN}$ & \multicolumn{2}{c}{Number Density $n$ (arcmin$^{-2}$)} & Pearson Correlation & \multicolumn{2}{c}{Best Fit Parameters} & $\chi^{2}_{dof}$ \\
& $(\circ)$ & $\times10^{5}$ & Detected & Useable & $\rho_{i}$ & $m_{i}$ & $c_{i}$ & \\
\hline 
\hline
\multirow{2}{*}{eMERLIN} & \multirow{2}{*}{90} & \multirow{2}{*}{4} & \multirow{2}{*}{128.1} & \multirow{2}{*}{63.6} & $0.702\pm0.007$ & $-0.180\pm0.012$ & $0.008\pm0.004$ & 1.281\\
& & & & & $0.706\pm0.007$ & $-0.158\pm0.012$ & $-0.001\pm0.004$ & $1.280$ \\ \hline

\multirow{2}{*}{eMERLIN} & \multirow{2}{*}{60} & \multirow{2}{*}{5} & \multirow{2}{*}{132.1} & \multirow{2}{*}{75.8} & $0.790\pm0.005$ & $-0.144\pm0.008$ & $0.001\pm0.003$ & 1.306 \\  & & & & & $0.799\pm0.005$ & $-0.128\pm0.008$ & $0.003\pm0.003$ & 1.062 \\ \hline

\multirow{2}{*}{LOFAR} & \multirow{2}{*}{90} & \multirow{2}{*}{2} & \multirow{2}{*}{121.4} & \multirow{2}{*}{50.0} & $0.687\pm0.009$ & $-0.085\pm0.013$ & $0.009\pm0.005$ & 0.843 \\
& & & & & $0.716\pm0.008$ & $-0.034\pm0.012$ & $-0.001\pm0.005$ & $1.291$ \\ \hline

\multirow{2}{*}{LOFAR} & \multirow{2}{*}{60} & \multirow{2}{*}{2} & \multirow{2}{*}{128.3} & \multirow{2}{*}{59.3} & $0.778\pm0.006$ & $-0.068\pm0.008$ & $0.010\pm0.004$ & 0.961 \\ & & & & & $0.816\pm0.005$ & $-0.075\pm0.007$ & $0.003\pm0.003$ & $1.060$ \\ \hline

\multirow{2}{*}{LOFAR (N)} & \multirow{2}{*}{60} & \multirow{2}{*}{2} & \multirow{2}{*}{127.1} & \multirow{2}{*}{52.6} & $0.759\pm0.007$ & $-0.084\pm0.010$ & $0.008\pm0.004$ & 1.270 \\ & & & & & $0.749\pm0.007$ &   $-0.099\pm0.011$ & $0.002\pm0.004$ & 0.708 \\ \hline\hline

\end{tabular}
\end{minipage}
\end{table*}

\subsection{Adding Noise}
The simulations above contain background fluctuations due to side-lobes, but do not contain measurement noise on the visibilities; we now consider the addition of this noise to the simulations. Within MeqTrees, the required input is the standard deviation of the Gaussian from which the noise for each visibility datum is drawn. For this test we restrict ourselves to the LOFAR DEC $=+60^{\circ}$ measurement set. 

\subsubsection{Assessing Noise Levels}

To assess the noise levels in the image plane corresponding to the noise input parameter in the MeqTrees software, we write the total noise $n_{t}$ in an image as
\begin{equation}
n_{t}^{2}=n_{sl}^{2}+n_{m}^{2},
\end{equation}
where $n_{sl}$ is noise associated with side lobes (which is present even in the absence of measurement noise) and $n_{m}$ is the contribution arising from the corruption of the visibilities. We use the noise-free simulations described in the previous section to estimate the $n_{sl}$ term, and we run test simulations with varying noise in the visiblilities to estimate the $n_{t}$ terms. Hence we can estimate the $n_{m}$ terms using the equation above. 

We find that the relationship between the measured $n_{m}$ and the visibility noise is linear, and calibrate the chosen level of noise accordingly. We calculate the SNR of each object in the image by  measuring the flux density in an aperture of 0.75$^{\prime\prime}$ around its known position, and dividing through by the total noise  $n_{t}$ integrated in the aperture.

\corr{We choose the noise level such that our objects have SNR $\simeq10$; this is the typical level to which weak lensing measurements are currently made.} We found that in MeqTrees parameter units, a choice of 20 for the visibility noise creates an image where the mean of the SNR distribution of the objects is $\langle\textrm{SNR}\rangle\simeq9.5$, and this is the value that we adopted. \corr{We note also that we have neglected to simulate the effect of primary beam attenuation which would increase the effective noise radially from the image centre. As we have not included this effect this amounts to assuming that we are in the beam centre where the effect is negligible.}

\begin{figure}
\centering
\includegraphics[width=\linewidth]{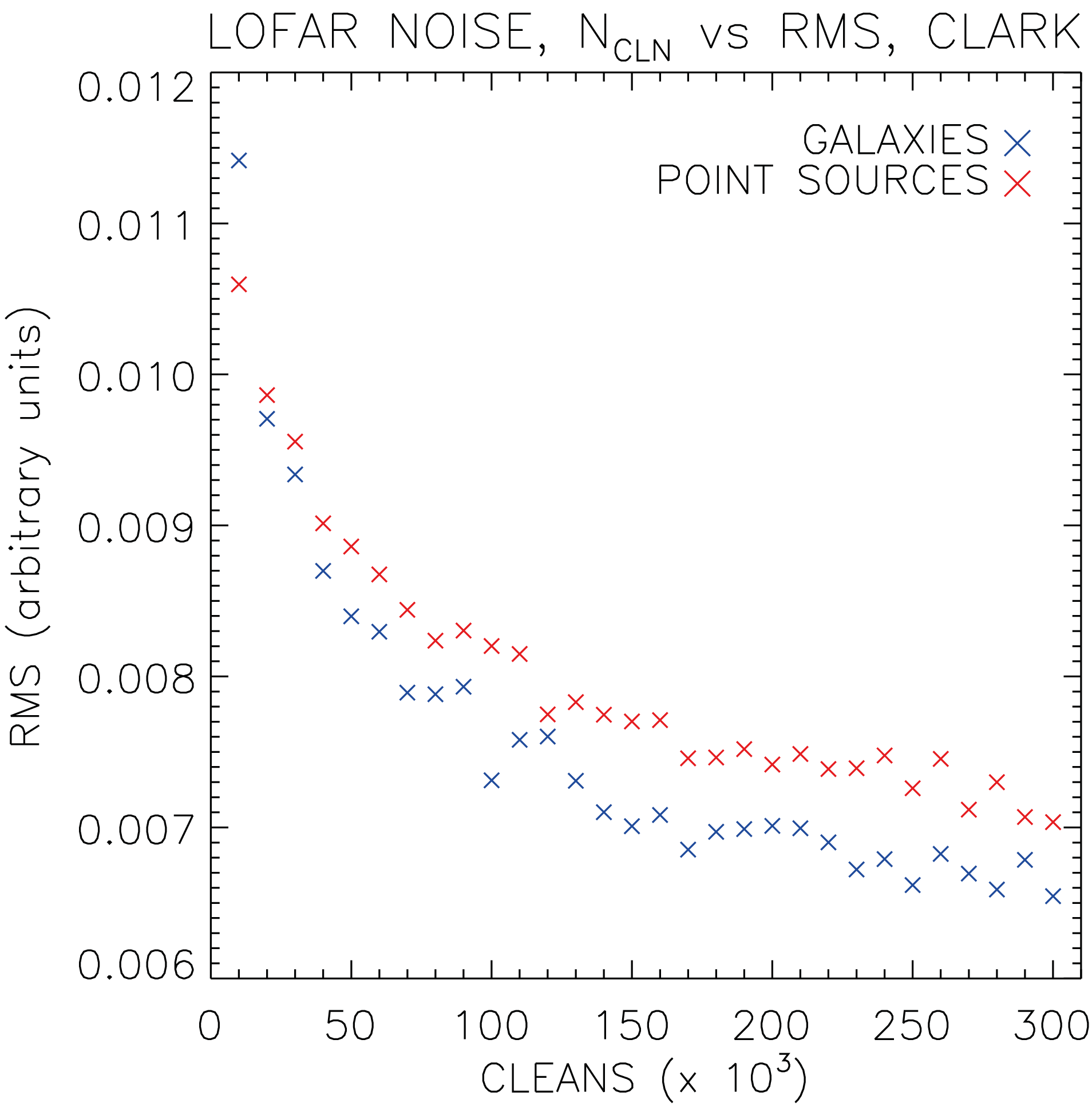}
\caption{Mean rms residuals for varying numbers of Clark CLEAN components for the LOFAR DEC $=+60^{\circ}$ case. \label{fig:noisecleans}}
\end{figure}

With this choice of visibility noise we carry out the same procedure as before. We first determine how many CLEAN components we need to sufficiently deconvolve the beam, by running a similar test to the noise free case, the result of which is shown in Figure \ref{fig:noisecleans}. 

We find that there is a different behaviour to that seen in the noise free case (c.f. Figure \ref{fig:clean}). We see that there is a reduction in the residuals as we increase the number of CLEAN components; at high numbers of components, both the point source and galaxy rms flattens out. In our noise free simulations we adopted a value of $1\times10^{5}$ and $2\times10^{5}$ CLEAN components for point sources and galaxies respectively for this MS. We see that with the addition of noise, point sources require more CLEANing than before, while for the galaxies $2\times10^{5}$ iterations still seems sufficient. We do not find a minimum in the rms residuals for either the point sources or the galaxies, but since after $N_{CLN}=2\times10^{5}$ there is only slow reduction, this is the value we have chosen. The choice will be vindicated if there is good correlation between measured and input ellipticities.

\begin{figure}
\centering
\includegraphics[width=\linewidth]{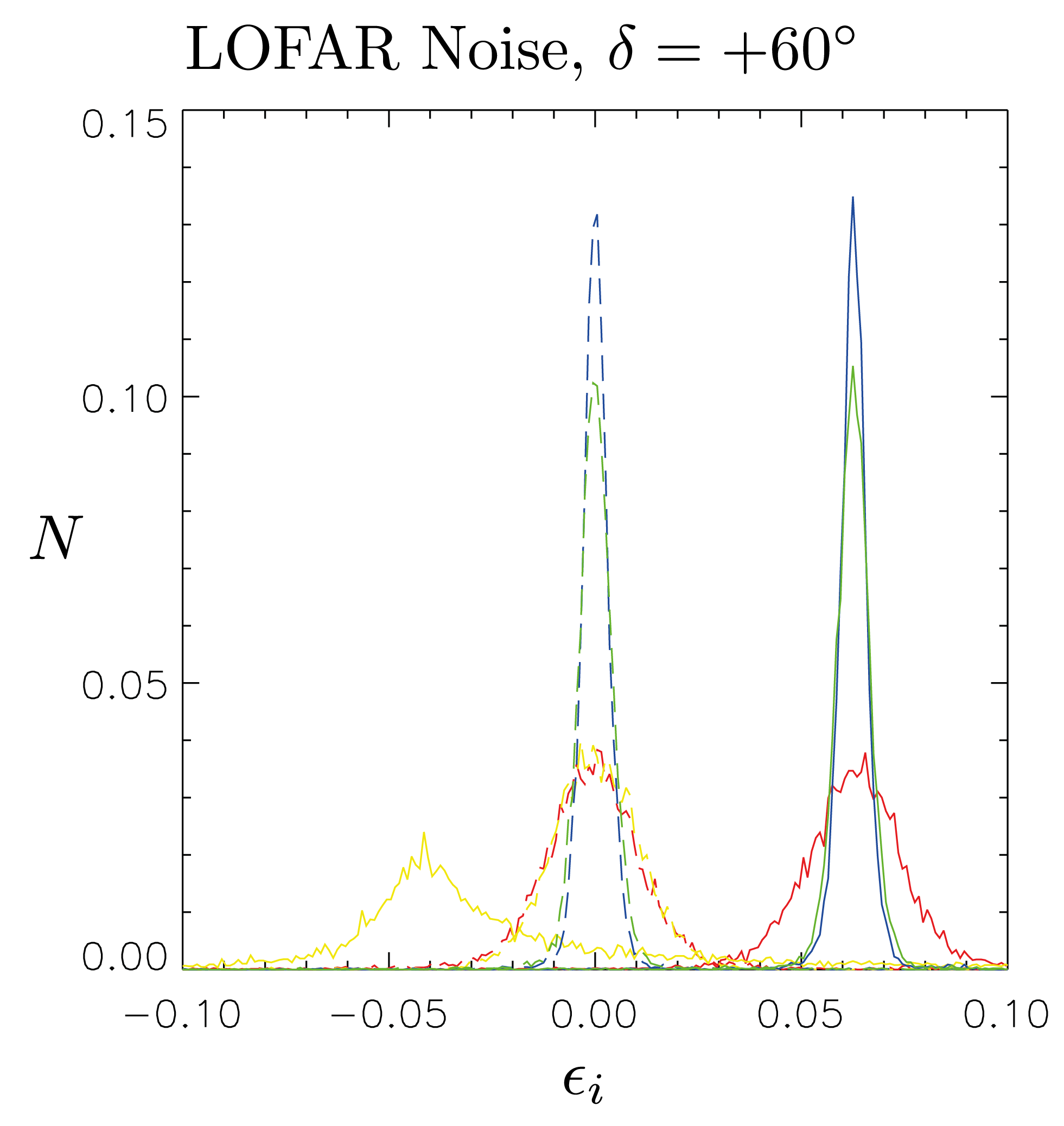}
\caption{Normalised point source ellipticity histograms for different choices of $\beta$, for the noisy LOFAR DEC $=+60^{\circ}$ case. Red curves correspond to the shapelet optimisation algorithm of \citep{2005MNRAS.363..197M}, and the yellow, blue and green curves correspond to decompositions done with fixed $n_{max}=10$ and $\beta=1, 1.5, 2$ pixels respectively. The solid lines show $\epsilon_{1}$ and the dashed show $\epsilon_{2}$. \label{fig:noisedelhists}}
\end{figure}

We now consider the noisy point source shape measurements. We perform a similar analysis for the shapelet modelling as in Section \ref{sec:point} to find a good choice for the $\beta$ parameter. We decompose all the sources using the shapelet optimisation algorithm as well as using our fixed $\beta$ and $n_{max}$ approach. Again, we use a fixed $n_{max}=10$ and use $\beta$ values of 1, 1.5 and 2 pixels. We show the  corresponding results in Figure \ref{fig:noisedelhists}. 

Immediately we note that the addition of noise has broadened out the ellipticity distributions in contrast to those measured in Figure \ref{fig:delhists}. As before, we find that fixing the $\beta$ parameter seems to produce a cleaner measurement than the standard shapelet optimisation algorithm, and we see again that a choice of $\beta=1.5$ pixels produces the most tightly peaked histogram. We again adopt this value for $\beta$ and an $n_{max}=10$ in our modelling of the point sources and the PSF. 

\begin{figure*}
\centering
\includegraphics[scale=0.5]{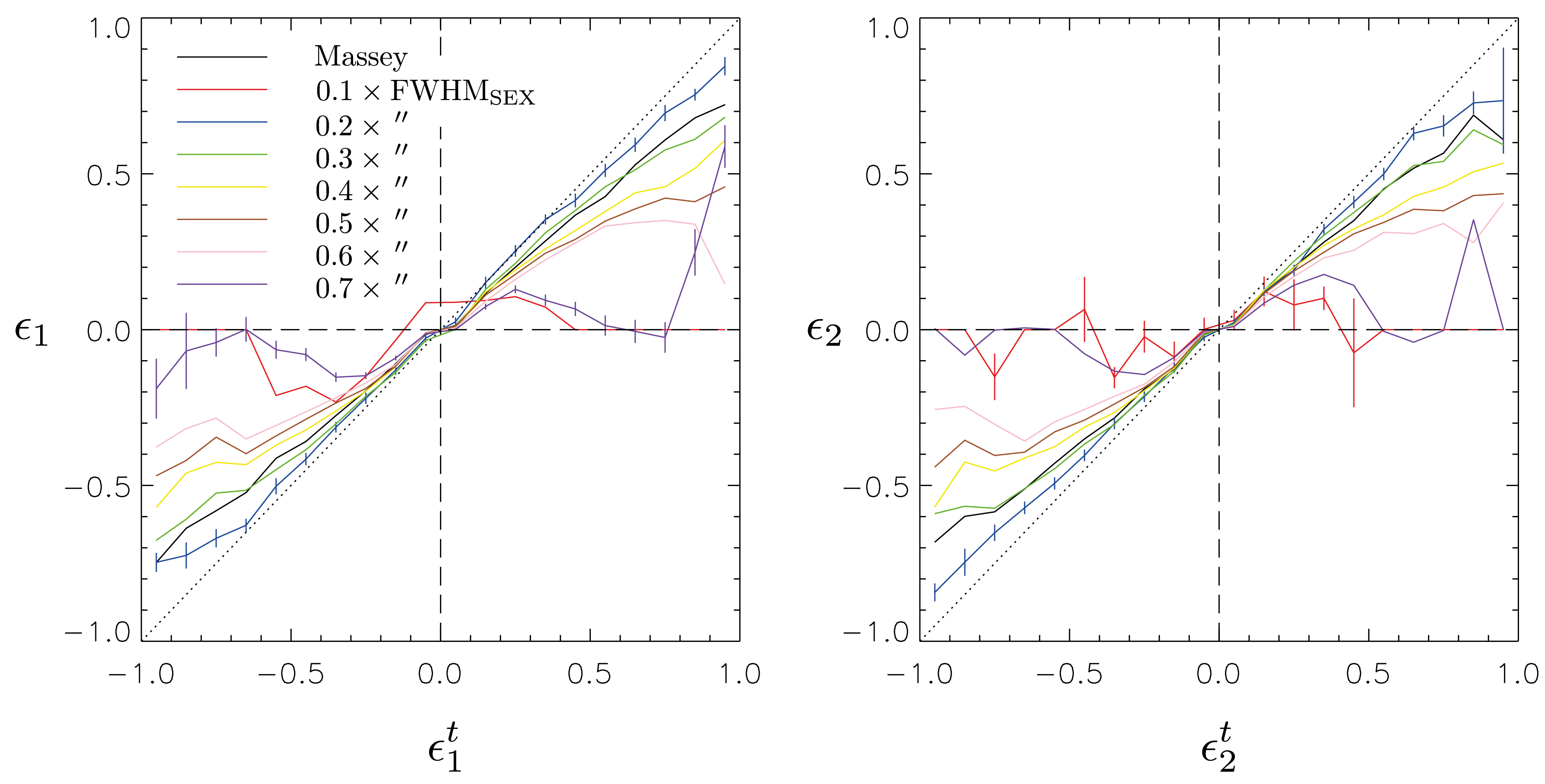}
\caption{Ellipticity comparisons for different choices of $\beta$ for the noisy LOFAR DEC $=+60^{\circ}$ MS. Black curves correspond to the shapelet optimisation algorithm of \citep{2005MNRAS.363..197M}, and the red, blue, green, yellow, brown, pink and purple curves correspond to decompositions done with fixed $n_{max}=10$ and $\beta=0.1, 0.2, 0.3, 0.4, 0.5, 0.6 \textrm{ and } 0.7 \times \textrm{SExtractor FWHM}$ respectively.\label{fig:noisebeta}}
\end{figure*}

\begin{figure*}
\centering
\includegraphics[width=0.5\linewidth]{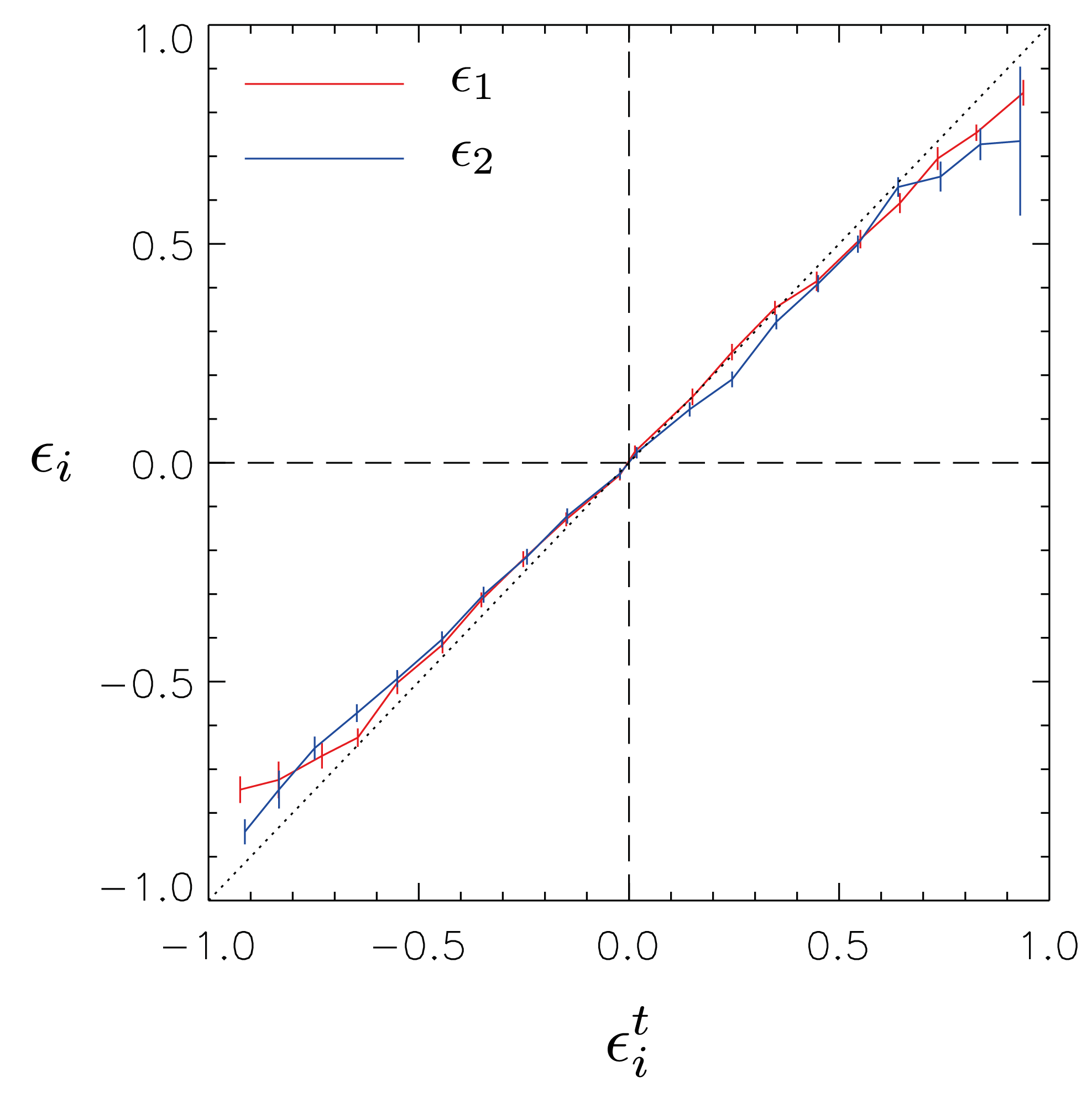}
\caption{Comparison of original and recovered ellipticities for the noisy LOFAR DEC $=+60^{\circ}$ MS.  \label{fig:noisegals}}
\end{figure*}

We use the estimated PSF to deconvolve the noisy galaxies. From the previous section, we expect that the $\beta$ parameter choice for the galaxies is  important, so we shapelet-decompose the galaxies with a variety of $\beta$ choices. As before we find a strong $\beta$ dependence when we examine how the measured ellipticities compare to the input ellipticities, as shown in Figure \ref{fig:noisebeta}, again we only show the smallest and largest error bars for clarity. Similarly to the noise-free case, we  find that fixing $\beta$ to 0.2 $\times \textrm{ the SExtractor FWHM}$ estimate yields the best results for $m$ when we compare the measured and input ellipticities. We show in Figure \ref{fig:noisegals} our final result of comparing the input and measured ellipticities in our noisy LOFAR DEC $=+60^{\circ}$ simulation, also summarised in Table \ref{tab:gals} as LOFAR (N). 

We find that the addition of the visibility noise has  degraded our slope measurement to $\epsilon_{1}-\epsilon_{1}^{t}=(-0.084\pm0.010)\epsilon_{1}^{t}+(0.008\pm0.004)$, and $\epsilon_{2}-\epsilon_{2}^{t}=(-0.099\pm0.011)\epsilon_{2}^{t}+(0.002\pm0.004)$. While this represents an increase in the calibration bias over the noise-free case, the calibration bias remains at a modest 10\% level.

\section{Conclusions and Future Work}
\label{discussion}

In this paper, we have made an exploratory study of a radio imaging pipeline suitable for studying weak gravitational lensing, and have tested a shear measurement pipeline which has been adapted for radio data. Using our simulations, we can obtain a better understanding of suitable shear measurement techniques to use with eMERLIN, LOFAR and ultimately the SKA. 

We have constructed a pipeline to simulate current and future weak lensing observations with radio interferometers. We were motivated by \citet{2010MNRAS.401.2572P} in which we attempted to measure a weak lensing signal using radio data from MERLIN and VLA. In that work we found systematic contamination in the data; indeed, a primary conclusion of that work was the need for a detailed study of the systematics involved in trying to measure weak lensing using radio datasets. In the current work, we have been able to assess some of the possible systematics, and have demonstrated the reliability of our shape measurement method on realistic simulated radio images.
Using the shapelets method we have created a set of images containing a collection of point sources and realistic galaxy shapes. The point sources were used to probe the behaviour of the beam, while the galaxies were used to test how well one can recover known ellipticities in the presence of a radio data reduction and imaging pipeline. The `true' images were run through our simulator to mimic observations made with the eMERLIN and LOFAR arrays, at two different declinations, and with different numbers of CLEAN iterations. 

We measured ellipticities for galaxies via a shapelets decomposition, including deconvolution of the PSF. As in our previous analysis, we have found that best results with shapelet decomposition were obtained by fixing the $n_{max}$ and $\beta$ parameters. The galaxy simulations showed a very strong $\beta$ dependance when we compared their true and measured ellipticities; we found that $\beta=0.2\times \textrm{SExtractor }$ FWHM gave the most faithful galaxy ellipticities. 

Given our best shapelet models, we were then able to compare the true and measured ellipticities in our catalogues. We found highly correlated results, with all four MSs having Pearson correlation coefficients close to one. All MSs showed no evidence for an additive bias to the ellipticity measurements, and showed a modest ($\simeq10$\%) multiplicative bias.

We added measurement noise to our simulations, fixing the visibility noise so that the resulting galaxies had a SNR distribution peaking at SNR $\simeq 10$. In this case, we found similar results for the shapelet ellipticity measurements as with the simulations only containing side-lobe noise; the multiplicative bias for ellipticity measurement remains at the 10\% level.
These results are encouraging since they suggest that we can already recover well the true shape of sources from radio data, with well-motivated choices for the number of CLEANs, and the shapelet scale size. Given further studies of  systematics effects, we hope to further reduce the calibration bias. 

In this work we have demonstrated the feasibility of making weak lensing measurements in the presence of realistic features of radio observations. Clearly, there are important extensions to what has been explored here: 

\begin{itemize}
\item \corr{In this study we have confined ourselves to small scale images, in order to CLEAN in a reasonably short time. The image size needs to be upscaled in order to probe position dependent PSF effects.}

\item We have restricted ourselves to one CLEAN algorithm; the performance of a range of deconvolution techniques should be tested.

\item We have used a uniform weighting scheme; weighting this way should maximise the contribution from the longer baselines, resulting in better resolution images. Natural weighting also exists, which give more sensitive images. Weak lensing is unique in that it requires both sensitivity and high angular resolution images; there are weighting schemes that try to find a compromise between these two criteria (e.g. Briggs weighting, \citealp{BRIGGS}). Assessing the shape measurement improvement/degradation between uniform, natural and intermediate weighting, and comparing this to any increase/decrease in source counts will be valuable.

\item In our construction of the MSs we have used particular frequency and time averaging configurations; these can dramatically increase/decrease the size of the MS.  Examining how different frequency and time averaging configurations affect shear estimates is a further important line of enquiry. \corr{We have also neglected all time and frequency smearing effects that also require further investigation.}

\item \corr{We have assumed calibration techniques are/will be good enough to perfectly remove the bright sources from the data.}

\item In this work we have used the shapelet method for shape measurement. The use of the shapelets method for radio data is well motivated, particularly by the Fourier invariance of the shapelet basis functions and the possibility of shape measurement directly in the $uv$ plane. However, in weak lensing, there are a considerable number of  methods for shape measurement, see for example \citet{2012MNRAS.423.3163K}. Exploration of radio weak lensing should include a study of how these existing methods fare with radio image simulations. 

\end{itemize}

In this analysis we have demonstrated an approach to weak lensing measurements at radio wavelengths;  encouragingly we have seen that our shape measurement methods and deconvolution techniques provide us with  shape measurements that are only modestly biased. We have pointed out several open questions that still remain to be answered in this field, some of which this pipeline should be able to answer. These simulations can be extended to simulate actual weak lensing fields, where the input images contain realistic lensing shear, and we can assess how well we  recover a cosmic signal. 

There are a range of new radio facilities that are/will be capable of producing data with which weak lensing measurements will be made. Understanding the relevant systematics, and knowing how well shape-measurement methods perform, is an important step towards using arrays such as a full Europe-wide LOFAR, and ultimately the SKA, for weak lensing.

\section*{Acknowledgements}
PP is funded by a NRF SKA Postdoctoral Fellowship. PP would like to thank Mathew Smith, Russell Johnston and Matt Jarvis for all the helpful discussions during the length of this project. BR acknowledges support from the European Research Council in the form of a Starting Grant with number 240672. FBA acknowledges the support of the Royal society via an RSURF. OMS's research supported by the South African Research Chairs Initiative of the Department of Science and Technology and National Research Foundation. Any opinion, findings and conclusions or recommendations expressed in this material are those of the authors and therefore the NRF and DST do not accept any liability with regard thereto.

%BIBLIOGRAPHY
\bibliographystyle{mn2e}

\label{lastpage}

\end{document}